\documentclass[aip,reprint]{revtex4-1}

\usepackage{graphicx}
\usepackage{dcolumn}
\usepackage{bm}
\usepackage{amsmath}
\usepackage{soul}
\usepackage{url}
\usepackage{hyperref}

\draft 

\begin{document}


\title{Deriving Reliable Nucleation Rates from Metadynamics Simulations: Application to Yukawa Fluids} 



\author{B.~Arnold}
\affiliation{Laboratory for Laser Energetics, University of Rochester, Rochester, New York 14623, USA}
\affiliation{Department of Physics and Astronomy, University of Rochester, Rochester, New York 14611, USA }

\author{J.~Daligault}
\author{D.~Saumon}
\affiliation{Los Alamos National Laboratory, P.O. Box 1663, Los Alamos, New Mexico 87545, USA}

\author{S. X.~Hu}
\email{shu@lle.rochester.edu}
\affiliation{Laboratory for Laser Energetics, University of Rochester, Rochester, New York 14623, USA}
\affiliation{Department of Physics and Astronomy, University of Rochester, Rochester, New York 14611, USA }
\affiliation{Department of Mechanical Engineering, University of Rochester, Rochester, New York 14611, USA }


\date{30 January 2026}

\begin{abstract}

In order to solidify the usefulness of metadynamics in studying nucleation of crystals from supercooled liquids, we provide a specific procedure to calculate nucleation free energy barriers. After a pedagogical review of the important elements of classical nucleation theory and how metadynamics is used to find nucleation free energy barriers, we explain the benefits of local collective variables over more common global collective variables. We show how a metadynamics free energy barrier must be carefully postprocessed so that classical nucleation theory can be applied to calculate nucleation rates. We apply our procedure to a Yukawa plasma and show that a particular physically-motivated fit to metadynamics data reproduces low-temperature reference data, justifying the usefulness of metadynamics to predict nucleation rates and the nucleation critical temperature.
\end{abstract}


\maketitle 

\section{Introduction}

When liquids are cooled below their freezing point, they form crystals. This process is ubiquitous and seemingly simple, but there is no single tool that can predict the mechanism, final crystal structure, and timescale of crystallization in all systems. The lack of a simple, universally-applicable method leaves many crystallization problems unsolved. This gap impedes our ability to understand the history of the earth’s core and magnetic field \cite{Huguet2018, wilson2023, Gao2025}, predict the structure of carbon in high pressure experiments \cite{Kraus2017, BC8, Shi2023}, and explain the cooling history of white dwarf stars \cite{Tremblay2019, Qbranch, blouinNeDist2021, Bedard2024}. Closing this gap will also help with  engineering of closer-to-home industrial processes used to create batteries and drugs \cite{Fu2022}.

Two advances are needed to understand crystal nucleation from the liquid in any system of interest. (1) A general framework must be constructed that can predict the mechanism by which crystals nucleate from the liquid. Great progress has been made in this area with increased understanding of two-step nucleation \cite{Lutsko2019, Iida2023, Alexandrov2023, Gispen2025}, the possible impact of polymorphs \cite{Leyssale2005, Bulutoglu2023, Gao2025, Gispen2025.2, Desgranges2025}, and the role of spontaneous fluctuations leading to prenucleation clusters or regions of local symmetry different from the bulk \cite{Ghiringhelli2007, Fu2022, Aslanov2025}. (2) Robust computational methods are needed to calculate the rate at which nucleation occurs along the known pathway for each particular system. Many methods have been used for this purpose, including classical density functional theory, brute force simulations, and numerous free energy-based enhanced sampling and path sampling methods, none of which is individually suitable to be applied to all interesting systems and conditions \cite{Sosso2016, blow2021, Finney2023}.

Metadynamics (metaD) is one of the free-energy based enhanced sampling methods that has been used before to compute nucleation rates. A goal of this work is to consolidate the applicability  of metadynamics to problem \#2. We focus on the Yukawa one-component plasma (YOCP) because it appears to nucleate via a classical pathway, which alleviates some concern about problem \#1. It is a widely relevant system because the one-component plasma is used as a simple model for charged colloids, dusty plsamas, electrolytes, and even extreme astophysical environments. We demonstrate that metaD can provide quantitatively accurate nucleation rates for the YOCP by comparing to brute force simulations and seeded simulations under identical conditions \cite{Arnold2025}.
Such comparisons between different computational methods are sparse in systems other than hard spheres \cite{Filion2010, Gispen2023}, and here we have the opportunity to explore whether the strength of the interaction between particles affect the consistency between methods.

This paper begins with a brief discussion of the YOCP model in section \ref{sec:sims}.  Section \ref{sec:nucleation_theory} covers the classical picture of nucleation, its limitations, and an extension. This is followed by a pedagogical review of metadynamics (Sec. \ref{sec:metaD_fundamentals}) and the the benefits of using local collective variables (Sec. \ref{sec:CVs}). In section \ref{sec:FES_interpretation}, we provide novel corrections that must be applied to metadynamics to theoretically justify using metadynamics with classical nucleation theory to calculate nucleation rates. Finally, in section \ref{sec:nuc_rates}, we use a physically-motivated fitting procedure to extract classical nucleation theory parameters from the metadynamics results to predict nucleation rates. We demonstrate our fitting method captures enough physical information from metaD simulations performed at one temperature to predict nucleation rates at other temperatures. The favorable comparison with brute force reference calculations in section \ref{sec:brute_force_nucleation} shows that these nucleation rates are reliable.

\section{Model and Simulations}
\label{sec:sims}

Our system of interest is the Yukawa one-component plasma (YOCP). This consists of a collection of $N_\text{tot}$ particles in a cubic box of volume $V$ under periodic boundary conditions, with a number density $n=N_\text{tot}/V$. These particles all have an identical mass $M$, charge $Q$, and interact via a screened potential
\begin{equation}
    V(r) = \frac{Q^2}{r} e^{-\kappa r/a}
\end{equation}
where $\kappa$ is the ratio of the average interparticle distance $a=\left(\frac{3}{4 \pi n}\right)^{1/3}$ to a screening length. For consistency with our prior work \cite{Arnold2025} and to explore a range of interaction lengths, we use $\kappa = 0, 2,$ and $5$. The screening length of the potential is inversely proportional to $\kappa$, so $\kappa=0$ corresponds to a pure Coulomb interaction while $\kappa=5$ gives a strongly screened, short range potential.

For ease of comparison between $\kappa$ values, we use nondimensional temperatures and times. Each value of $\kappa$ has a different melt temperature $T_\text{m}$, known from free energy calculations \cite{Farouki1994, HamaguchiTriplePoint1997}, which sets the temperature scale, so we use the reduced temperature $\Theta = T/T_\text{m}$. Our simulations all occur between $\Theta = 0.6$ and $0.85$, so the temperature is below the melt temperature and the solid phase is thermodynamically preferred. Similarly, since particle motions occur on a timescale set by thermal velocity, defined as $v_{th} = \sqrt{2k_BT/M}$, we use the characteristic time scale $\tau = a/v_{th}$ throughout this work. 

All molecular dynamics (MD) simulations were performed with the open-source LAMMPS simulation package \cite{LAMMPS}, and the metadynamics biasing was implemented with the PLUMED package, which is integrated with LAMMPS to provide tools for enhanced sampling simulations \cite{PLUMED}. The equations of motion are integrated with a time step of $\tau/100$, which is sufficiently small that numerical error does not cause drifts in total energy. A Nose-Hoover thermostat maintains the desired temperature in each simulation.

\section{Classical Nucleation Theory: Assumptions, Deviations, and Effective Parameters}
\label{sec:nucleation_theory}

When a liquid is moderately supercooled, forming a crystal is thermodynamically favorable. However, the liquid is usually metastable with respect to the solid, meaning that some finite time is required before the solid forms. Qualitatively, this occurs because the bulk solid can not grow until there is a small crystalline cluster, or nucleus, onto which additional particles can attach. When there are no impurities to jump-start this process, a cluster has to overcome a free energy barrier to fluctuate (nucleate) into existence. This fluctuation takes a finite time to occur, limiting the rate at which solid regions can form in the bulk liquid.

\subsection{The Classical Picture}
\label{sec:CNT}

\begin{figure}
    \centering
    \includegraphics[width=1\linewidth]{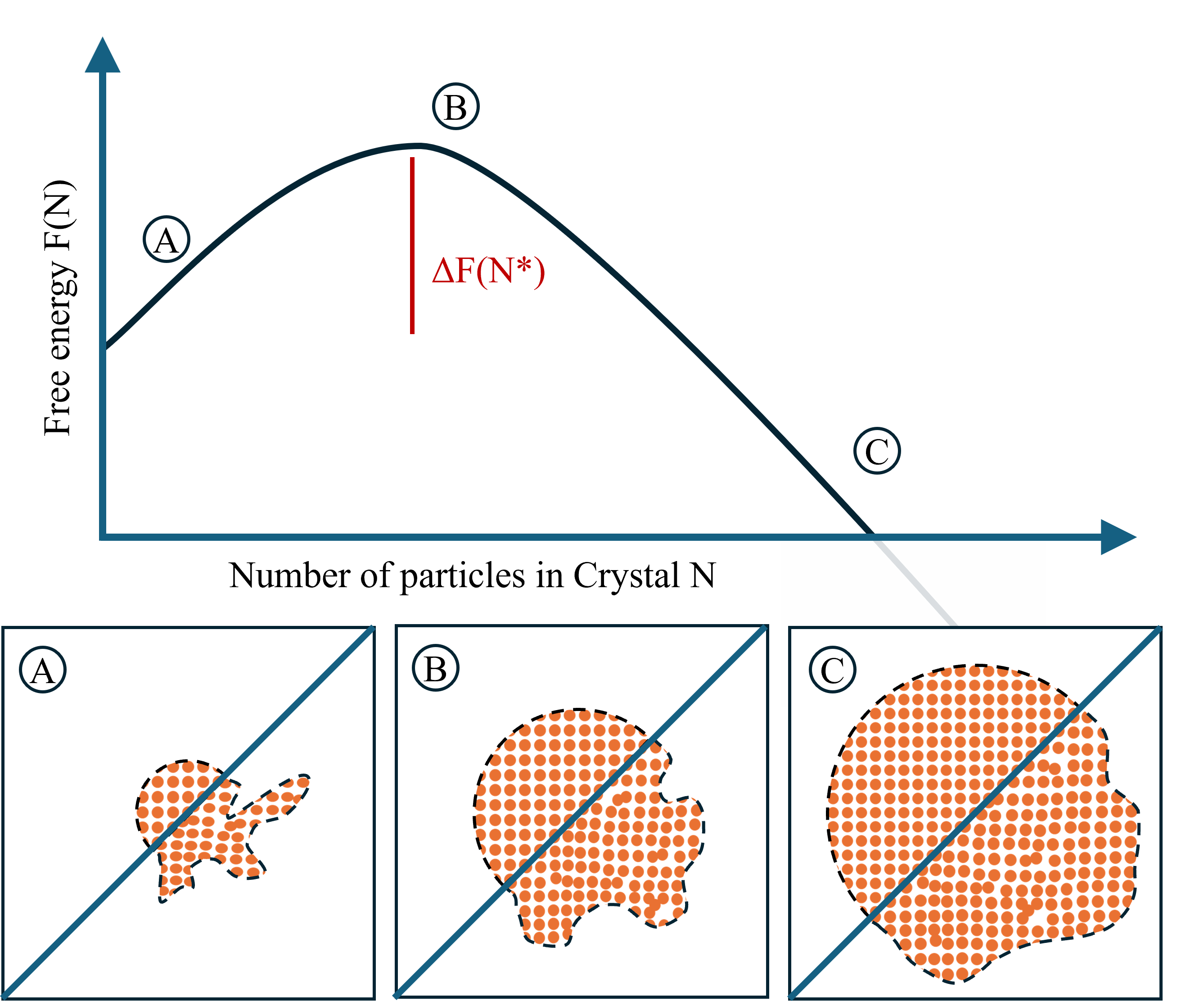}
    \caption{(Upper panel) CNT free energy surface [Equation \eqref{eq:freeEnergy}]. (Lower panels) Differences between idealized spherical CNT clusters (upper left half of panels) and more realistic clusters (lower right half of panels)}
    \label{fig:CNT_FES}
\end{figure}

The classical nucleation theory (CNT) is the simplest quantitative description of nucleation\cite{Kelton1991}. It supposes that the Helmholtz \footnote{When a small solid cluster forms in a large liquid region, the Gibbs and Helmholtz free energies are the same. Since we use NVT simulations, we discuss specifically the Helmholtz free energy.} free energy input required for a single spherical crystal to reach a size $N$, while surrounded by a homogeneous liquid, is
\begin{equation}
    \label{eq:freeEnergy}
    \Delta F(N) = \Delta \mu N  + \gamma N^{2/3}.
\end{equation}
Here, $\Delta \mu$ is the free energy difference per particle between the solid and liquid at constant pressure. In supercooled liquids, $\Delta \mu$ is negative, so the first term represents the free energy decrease associated with formation of the bulk solid. This is the driving force towards solidification. The second term is the surface contribution -- a positive free energy cost to forming an interface between a growing solid cluster and the surrounding liquid. $N^{2/3}$ is proportional to the surface area of a spherical cluster, so $\gamma$ is proportional to an energy per surface area. The CNT free energy \eqref{eq:freeEnergy} grows with $N$ until it peaks at the critical cluster size $N^* = \left|2 \gamma/3  \Delta \mu\right|^3$. Figure \ref{fig:CNT_FES} shows this schematically. Panel A shows a small cluster where growth is thermodynamically unfavorable due to the surface energy, B shows a cluster with $N\sim N^*$,  and C shows a large cluster where further growth is favorable because the bulk free energy contribution dominates.

The nucleation rate $J$ is defined as the rate at which a cluster can climb over the free energy barrier, reaching a size $N>N^*$ where it can grow to become a macroscopic crystal. CNT assumes that a growing crystal climbs the barrier stochastically, losing and gaining single constituent particles in discrete events that occur with rate $D_+$. This gives the CNT nucleation rate 
\begin{equation}
    \label{eq:nucRate}
    J = D_+ \sqrt{\frac{\left| \Delta \mu \right|}{6 \pi k_B T N^*}} \exp{\left\{-\Delta F\left(N^*\right)/k_BT\right\}},
\end{equation}
which depends on the height of the free energy barrier $\Delta F(N^*)$; An increased free energy barrier height exponentially suppresses the rate of nucleation. The prefactor to the exponential contains all the information about the kinetics of the transition \cite{Kelton1991}.

CNT is simple to use, but it includes many assumptions, any of which could be violated. If CNT is applied carelessly to a system that violates its assumptions, it can predict nucleation rates that are incorrect by many orders of magnitude. The assumptions are violated by multi-step pathways involving intermediate phases between the solid and liquid, pre-ordering of the liquid, or fusion of nuclei. \cite{Sosso2016, blow2021}. These violations require modifications to CNT to obtain correct nucleation rates (See e.g. refs. \onlinecite{Iida2023,Bulutoglu2023,Gispen2025}), but the YOCP studied here undergoes single-step nucleation.

Even when nucleation proceeds directly from the liquid to the final solid as predicted by CNT, the forming crystal always deviates somewhat from the CNT idealization. In writing Equation \eqref{eq:freeEnergy}, we assume that the forming crystal has a chemical potential equal to that of the bulk solid, that the surface is smooth and spherical while maintaining the thermodynamic properties of a planar solid-liquid interface, and that the surface energy is well-defined. In practice, however, there are stresses and defects in the crystal that modify its chemical potential \cite{deJager2024}, clusters may be nonspherical with diffuse solid-liquid interfaces \cite{Moroni2005, Leyssale2007}, and the cluster surfaces differ from a planar interface between a bulk solid and a bulk liquid \cite{Auer2001}.

Figure \ref{fig:CNT_FES} illustrates these problems. Panels A, B, and C contrast CNT-assumed crystallites with more realistic cases. The CNT assumptions are more closely obeyed by large, and especially postcritical ($N>>N^*$, as in panel C) clusters, but small clusters may be generated by rapid fluctuations \cite{Trudu2006}, or lack a distinct surface and ordered interior \cite{Moroni2005}. For these reasons, some authors suggest that CNT should not be used to calculate quantitative nucleation rates at all, but rather just to examine trends in the rates as parameters of the system like temperature vary \cite{Leyssale2007}.

\subsection{Effective CNT Quantities}

Traditionally, a CNT calculation uses $\Delta \mu$ for a bulk, defect-free solid and liquid, and $\gamma$ for a planar interface between a liquid and a perfect crystal. The calculation of $\gamma$ by thermodynamic integration, metadynamics, or the cleaving method is challenging \cite{Cheng2015, Yamamura2025, Ghoufi2025}, and they lead to nucleation rates differing from experiments and other methods by many orders of magnitude, and the claim that CNT does not predict accurate nucleation rates.

Expensive biased enhanced sampling, or unbiased path sampling simulations can predict nucleation rates without CNT \cite{Finney2023}. Unfortunately, these methods are complicated, and usually limited to a small range of temperatures. For all methods, reaching temperatures close to the melt temperature is prohibitive because the critical cluster becomes too large to fit in a simulation box. 

While it is not as accurate as other methods, CNT is computationally cheap and predicts nucleation rates at all temperatures. The accuracy of CNT can be rescued while retaining its benefits by using ``effective'' rather than ``real'' quantities. While CNT with the free energy surface (FES) as written in Equation \eqref{eq:freeEnergy} is inaccurate, one can calculate free energy barriers directly at a few conditions and fit parameters $\Delta \mu$ and $\gamma$ of a CNT-like free energy to these results. These are ``effective'' parameters because they are not the idealized bulk quantities assumed by CNT, but they capture unknown effects that cause deviations from CNT. This strategy produces accurate nucleation rates from only a few free energy calculations and maintains CNT's extensibility to any temperature \footnote{This extensibility only applies in regions of parameter space where the assumptions of CNT are reasonable. We show later that this means that results can be extended to higher temperatures.}.

Seeded simulations or biased enhanced sampling methods like metaD or umbrella sampling can provide the free energies for fitting. We previously used seeded simulations \cite{Arnold2025}, but this method can be inaccurate because it relies of artificial clusters rather than allowing simulations to find the lowest free energy structures \cite{espinosa2016} and because ambiguities in the identification of solid particles prevent accurate barrier height calculations \cite{Zimmerman2018}. MetaD or umbrella sampling are more accurate because they naturally allow diffuse, fractal-like, or nonspherical clusters \cite{Trudu2006}.

Prior studies have shown many times that effective CNT parameters are required to describe nucleation free energy barriers in Lennard-Jones systems \cite{bai2006, bai2008, Tipeev2018, cheng2017.2, baidakov2012, tenWolde1996} but it has rarely been applied to other systems like hard spheres \cite{Auer2001}, methane hydrate \cite{Arjun2023}, and water \cite{Lin2024}. 
We now extend this approach by applying the analysis to a new system and by demonstrating a fitting strategy for the effective CNT parameters that allows extrapolation to higher temperatures and reproduces nucleation rates calculated by brute force simulations at lower temperature.

\section{Metadynamics: Fundamentals, Collective Variables and Interpretation for Nucleation }
\label{sec:metaD}

Subsections A-C of this section include a review of the fundamentals of metadynamics, make explicit the previously-known benefits of using local collective variables, and provide novel corrections to the free energy surface required for consistency with CNT. These subsections are all educational, explaining why we used our particular metaD procedure to calculate nucleation free energy surfaces. Skipping to subsection \ref{sec:procedure} will simply provide the final procedure.

\subsection{Metadynamics Fundamentals}
\label{sec:metaD_fundamentals}
MetaD was initially developed to study free energy surfaces for biochemical applications \cite{Barducci2011, magsumov2024}, but it is now frequently used for studying transitions between different metastable solid phases \cite{Badin2021, Martonak2025},  or between a supercooled liquid and a nucleating solid \cite{Quiggley2009}.

Consider an unbiased molecular dynamics simulation of a supercooled liquid governed by the Hamiltonian $H_0$. This simulation will use $H_0$ to evolve all atomic coordinates $X(t)$ in time. Further consider a function $S(X)$, a function of atomic coordinates that measures crystallinity: it takes large values when $X$ is a crystal arrangement and small values when $X$ is disordered. $S(X)$ may be called an order parameter or a collective variable (CV) because it compresses complex information about the structure of $X$ into a single number. 

With unlimited computing power, one could use straightforward MD to evaluate free energy as a function of $S$. A histogram of values of $S(X(t))$ visited at time $t$ during a long simulation approximates the probability $P(S)$ of the system visiting a particular value of $S$, which is related to the free energy via

\begin{align}
    \label{eq:unbiased_FES}
    F(S)  &= -k_BT\ln\big(P(S)\big) \\
        &=  -k_BT\ln \left(\int dt ~\delta\Big[S-S\big(X(t)\big)\Big]\right),
\end{align} 
where $\delta$ is the Dirac delta. This free energy captures the likelihood of an unbiased system reaching some value of $S$; when $P(S)$ is small, $F(S)$ is large, and that value of $S$ will be rarely visited. Reasonable unbiased MD simulations do not visit large-$F(S)$ states. From such a simulation, it is obvious where $F(S)$ is large, but there is no way to calculate just how large it is.

To overcome this sampling limitation, metaD adds a bias potential $V_\text{b}(S)$ to the Hamiltonian that drives the simulations to sample otherwise-inaccessible configurations with high free energy. We then use the same idea underlying Equation \eqref{eq:unbiased_FES} to calculate the unbiased free energy\cite{Barducci2011}. 

\begin{figure}
    \centering
    \includegraphics[width=1\linewidth]{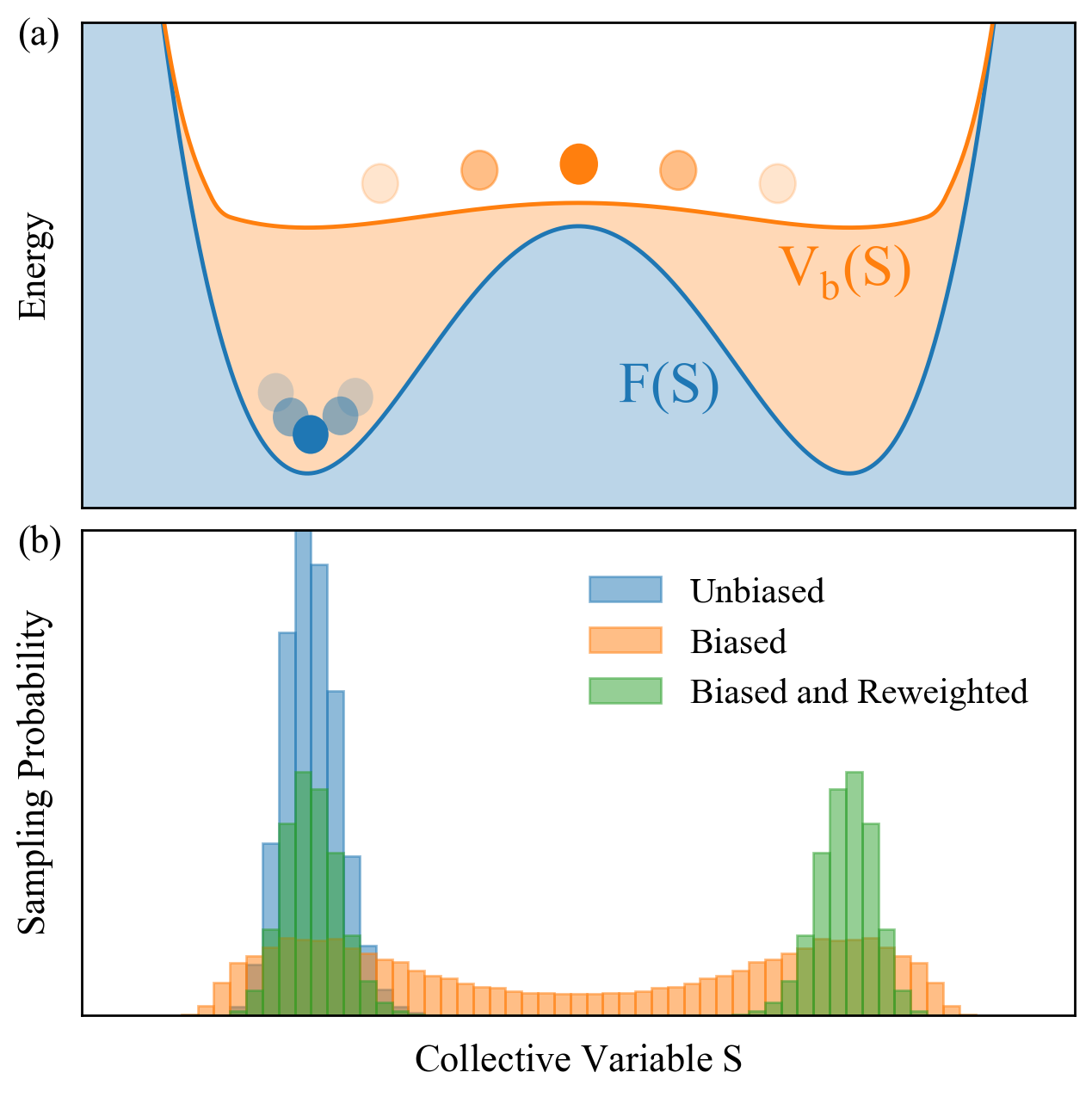}
    \caption{Illustration of the principles of metadynamics. (a) In blue, the system is stuck in a well of the unbiased free energy surface $F(S)$. With a compensating bias applied, the total free energy surface $F(S)+V_b(S)$ is nearly flat and the (orange) biased system can traverse anywhere. (b) The unbiased system (blue) reaches only a small range of $S$ values. When the bias is applied (orange) it can sample more $S$ values with similar probabilities. When the biased sampling is reweighted (green), the ``true'' sampling probability (corresponding to the result of an infinitely long unbiased simulation) is returned.}
    \label{fig:metaD_cartoon}
\end{figure}
Roughly, if a modified Hamiltonian $H_{\text{mD}} = H_0 +V_\text{b}(S)$ generates dynamics that sample all values of $S$, then the free energy surface can be calculated as
\begin{equation}
    F(S) = -k_BT \ln \left( \int dt~ \delta\Big[S-S\big(X(t)\big)\Big]e^{V_\text{b}(S(X(t)))/k_BT} \right)
\end{equation}
This expression is similar to the one used to calculate free energy from unbiased MD: it is a logarithm of a histogram accumulated during a simulation. However, since the biased simulation was pushed away from configurations with large $V_\text{b}(S)$, the factor of $e^{V_\text{b}(S(X))/k_BT}$ \emph{reweights} the histogram to restore probabilities that would be calculated by an infinitely long unbiased simulation\cite{Bonomi2009}. Note that we can also reweight to calculate the free energy surface with respect to some other variable that was not biased, like the number of solid particles in a cluster,  $N$.
\begin{equation}
\label{eq:reweight}
    F(N) = -k_BT \ln \left( \int dt ~\delta \Big[N-N\big(X(t)\big)\Big]e^{V_\text{b}(S(X(t)))/k_BT} \right)
\end{equation}

In our analysis, we use balanced exponential reweighting to estimate the free energy surface \cite{Schafer2020}. This is the same as the above in spirit, but it accounts for the fact that $V_\text{b}$ varies with time, which provides faster convergence of $F(N)$.

The form of $V_\text{b}(S)$ that compensates the intrinsic free energy surface $F(S)$ is unknown \emph{a priori}, so metaD provides a prescription to iteratively build $V_\text{b}(S)$ as a sum of many Gaussians. If $S$ is chosen well so that $V_\text{b}(S)$ causes sufficient sampling of all important configurations, metaD will provide accurate free energy surfaces. Making the appropriate choice of order parameters is one of the major challenges of metaD \cite{Bussi2020}.

\subsection{Collective Variables}
\label{sec:CVs}

MetaD is only effective when collective variables are carefully chosen to differentiate the metastable states from each other, and to differentiate the metastable states from all relevant transition states\cite{Bussi2020}. For our case, this means that $S(X)$ must be chosen so that it takes different values in the solid, the liquid, and in every step of the process by which the liquid transforms to the solid. In this section, we discuss why local CVs distinguish transition states between the liquid and solid better than commonly-used global CVs. We define the particular CVs used in this study to achieve adequate sampling of relevant transition states.

For systems with density or coordination differences between relevant states, simple CVs like potential energy, coordination number, or unit cell vectors are effective\cite{Santos2022, martonak2003, Badin2021}. We found that for the YOCP, energies and coordination numbers fluctuate within a bulk phase and differ little between the liquid and solid, making them unsuitable to drive the relevant transition. In this situation, symmetry-based CVs can be used for metaD studies of crystals forming from melt \cite{Giberti2015}. One such CV is the neighbor-averaged Steinhardt bond order parameter $Q_6(i)$, where i denotes a particle index. Reference [\onlinecite{Lechner2008}] describes how $Q_6(i)$ uses sixth order spherical harmonics to quantify the symmetry of the arrangement of neighbors around particle $i$. It takes a large value around 0.8 in BCC YOCP crystals, which are highly symmetric, and a low values around 0.2 in the YOCP liquid, which is disordered.  These CVs are implemented in PLUMED so they can be used to bias LAMMPS simulations \cite{Tribello2017, LAMMPS}.

We contrast using $Q_\text{tot}$, a global order parameter against using $Q_\text{in}$, a local order parameter, both based on $Q_6$. We show that $Q_\text{in}$ is more effective due to its ability to sample physically correct transition states. 

\subsubsection{Global CVs}

\begin{figure}
    \centering
    \includegraphics[width=1\linewidth]{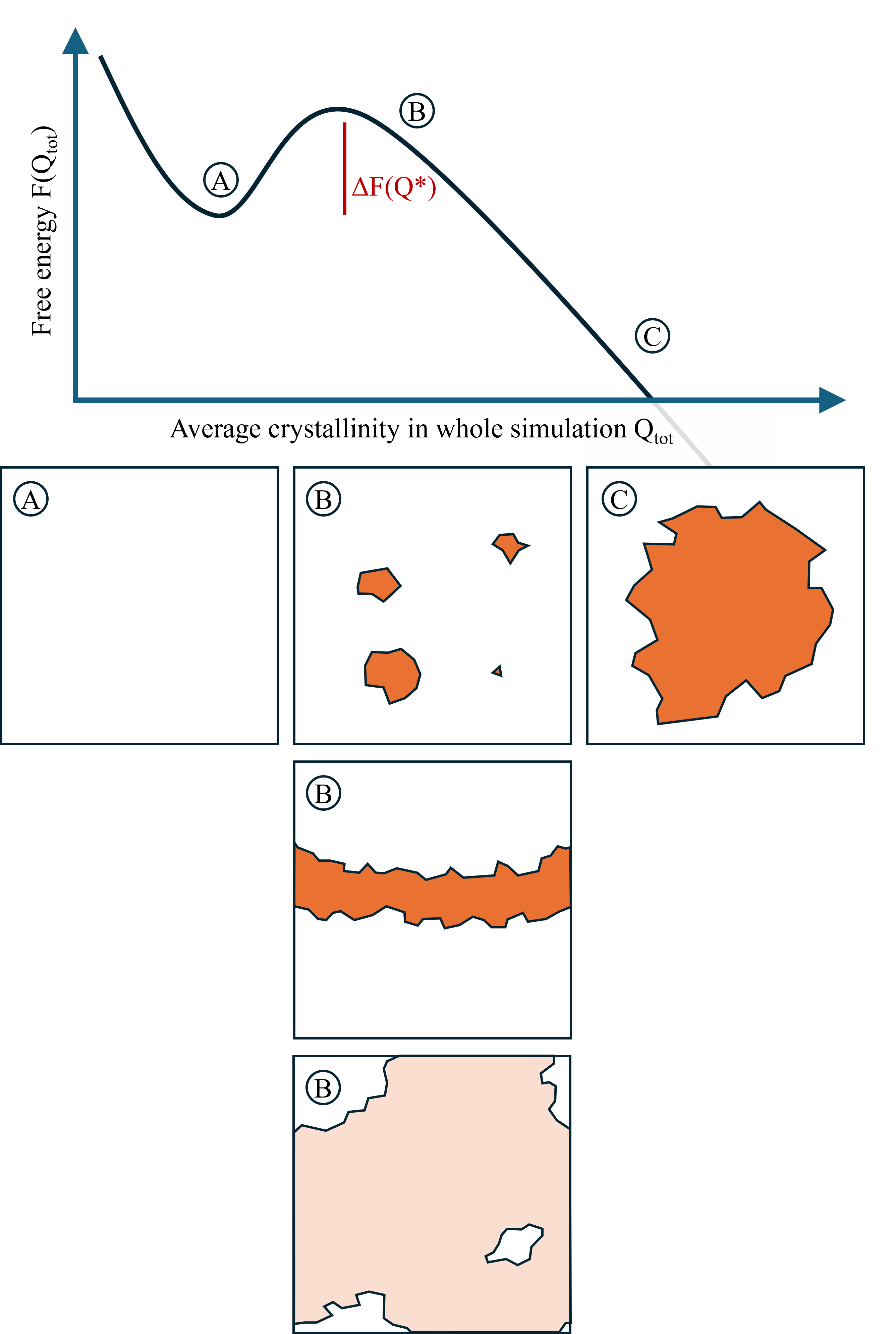}
    \caption{A schematic of the free energy surface generated by the globally-biased metadynamics method. (A) Configurations at the bottom of the liquid free energy well contain no large crystal clusters. (B) Near the top of the free energy barrier, the simulation generates multiple nuclei or other erroneous transition states. (C) The free energy becomes much lower than the liquid free energy in highly crystallized states because the crystal is thermodynamically favored over the liquid.}
    \label{fig:fullspace}
\end{figure}

When studying the transition between two phases, a common approach is to bias a global order parameter, like $Q_6$ averaged over all particles in the simulation:
\begin{equation}
    Q_\text{tot} = \frac{1}{N_\text{tot}}\sum_i^{N_\text{tot}} Q_6(i).
\end{equation} Fig. \ref{fig:fullspace} shows schematically the free energy surface $F(Q_\text{tot})$ and representative configurations generated by biasing $Q_\text{tot}$.  

This approach is simple. It yields a free energy surface with a small well around $Q_\text{tot}=0.2$ containing the metastable liquid. A second, much deeper well around $Q_\text{tot}=0.8$ contains the thermodynamically favorable solid state. A free energy barrier between the two wells prevents instantaneous nucleation of the solid from the metastable supercooled liquid.

A major challenge of this method is that the transition state is uncontrolled and prone to finite-size effects. To resolve the nucleation free energy barrier we are pursuing, our metaD simulations must generate the transition state at the top of the barrier: configurations containing a single critical cluster surrounded by liquid. The B panels of Fig. \ref{fig:fullspace} show that transition states at intermediate values of $Q_\text{tot}$ may not have a single large crystalline cluster. There may be multiple clusters near the critical cluster size, meaning that even if the simulation is driven to the value of $Q_\text{tot}$ where $F(Q_\text{tot})$ is maximized, the critical clusters of physical interest may not actually be sampled \cite{Leyssale2005, Quiggley2009}. This method could also sample other nucleation pathways that only exist in finite-sized simulations, like elongated critical nuclei reaching across periodic boundaries \cite{Quiggley2009} or global coordinated transformation where the whole system solidifies simultaneously \cite{Badin2021}. If, for any of these reasons, the simulation fails to generate configurations with a single nucleus unaffected by finite size effects, the method does not converge to a free energy surface applicable to large systems.

\subsubsection{Local CVs}
To overcome the issues with global biasing, one can instead bias only a subset of particles. We choose one particle as the ``central particle" and a radius $r_\text{b}$. All particles within $r_\text{b}$ of the central particles are ``inside" particles. There are $N_\text{in}$ of these, and the rest are ``outside" \footnote{This distinction is not actually binary. For numerical reasons, all CVs in metaD should be continuous, we use a switching function \cite{Tribello2017}}. This leads to two order parameters
\begin{align}
    Q_\text{in} &= \frac{1}{N_\text{in}}\sum_{i \in \text{inside}} Q_6(i) \\
    Q_\text{out} &= \frac{1}{N_\text{tot}-N_\text{in}}\sum_{i \notin \text{inside}}Q_6(i).
\end{align}
We bias $Q_\text{in}$ to drive nucleation while restraining $Q_\text{out}$ to ensure most of the simulation volume remains liquid (See supplementary material for details on the restraint on $Q_\text{out}$).

\begin{figure}
    \centering
    \includegraphics[width=1\linewidth]{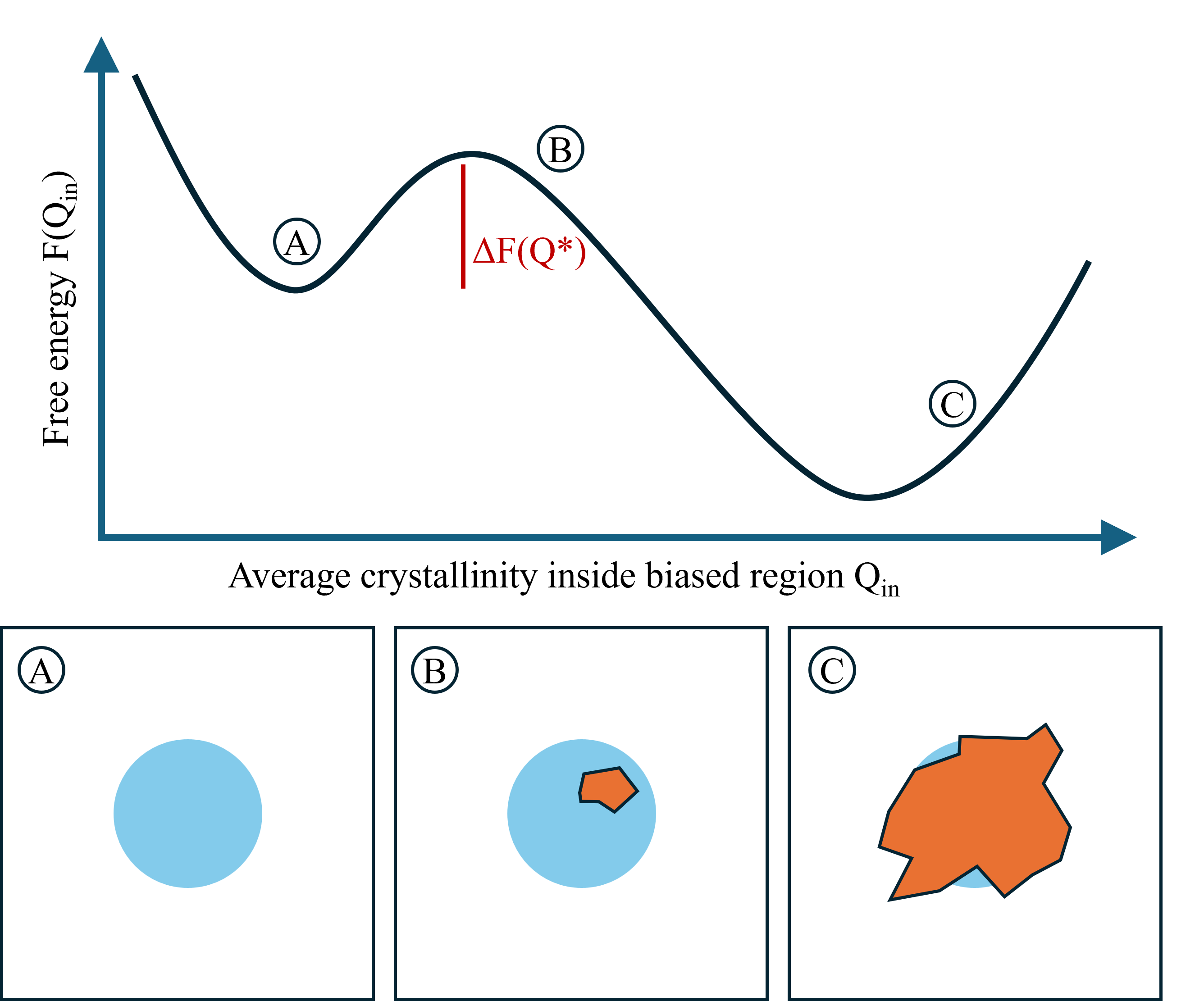}
    \caption{Cartoon of the free energy surface generated by the local biasing method. The metadynamics bias generates crystals in the light blue region, while the ``outside'' white region of the simulation is restrained to remain liquid. (A) There are no solid clusters at the bottom of the liquid free energy well. (B) The classical nucleation pathway is respected with only one cluster forming inside the biased region at the top of the free energy barrier. (C) The crystal grows to fill the ``inside'' region, but can't become larger to fill the whole simulation volume.}
    \label{fig:subspace}
\end{figure}
Figure \ref{fig:subspace} illustrates this local biasing strategy, which samples the classical nucleation path better than global biasing. Since most of the simulation is forced to remain liquid, it does not generate configurations with many near-critical nuclei. And, since the biased region is compact, the simulation will not generate nonphysical elongated nuclei or undergo coordinated phase transformation. For this reason, we use the local CV $Q_\text{in}$ to drive metadynamics simulations for calculating nucleation rates. 

However, the local biasing method fails at very low temperatures. Under these conditions, crystals containing several hundred particles form in the outside ``liquid'' region despite attempting to force it to remain liquid (see supplementary material). The simulation is unable to generate configurations with no clusters, so adequately sampling small cluster sizes becomes impossible. 

\subsubsection{CV comparison}

The tradeoffs between biasing global and local CVs mean that local biasing is practical at moderate temperature where it is possible to maintain a liquid region. At lower temperatures neither method is ideal, but only global biasing is possible. For the YOCP, local biasing only works at or above reduced temperatures $\Theta = 0.725$ where the free energy barrier is greater than $\sim 10 k_BT$.

\begin{figure}
    \centering
    \includegraphics[width=1\linewidth]{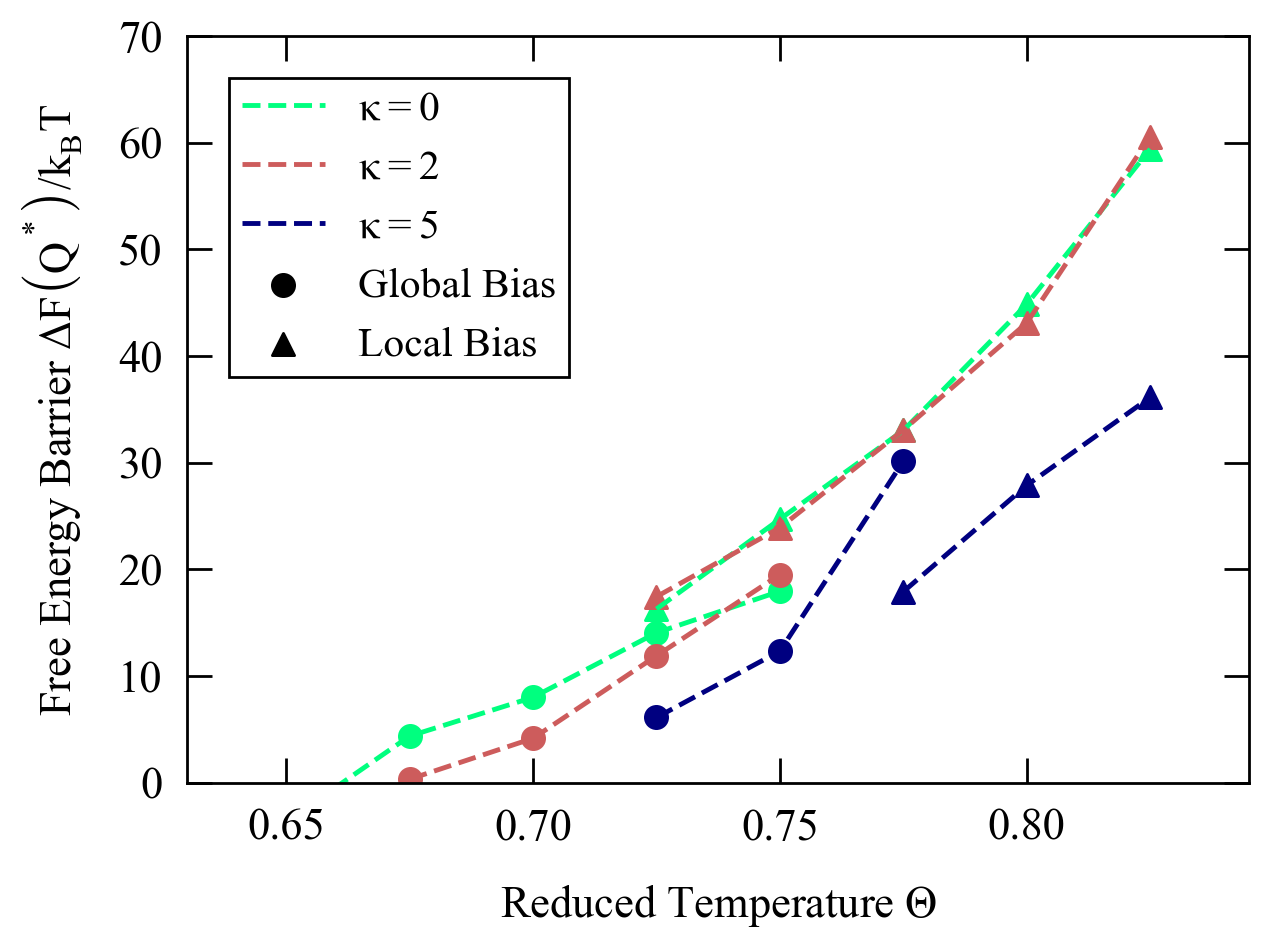}
    \caption{Nucleation free energy barrier heights for Yukawa plasmas with screening parameters $\kappa=0,2,$ and $5$. Triangles show the barrier heights calculated using local biasing at moderate temperatures, and circles show barriers from global biasing at lower temperatures.}
    \label{fig:DeltaF}
\end{figure}

We have used both of these methods in their respective region of applicability to calculate nucleation free energies. The free energies are functions of the biased order parameter ($Q_\text{tot}$ or $Q_{in}$) and they take on a maximum value at a location that we call $Q^*$. Figure \ref{fig:DeltaF} shows the height of the free energy barrier separating the liquid from the solid wells, $\Delta F\left(Q^*\right)$, from $\Theta = 0.65$ to $0.825$ from both biasing methods. This comparison demonstrates the correspondence between globally- and locally-biased metaD,
but we show in the next section that $F(Q)$ is not equivalent to the CNT $F(N)$ and should therefore not be used to predict nucleation rates. 

Our simulations show that the $F(Q_\text{tot})$ free energy barrier from globally-biased metaD vanishes around $\Theta = 0.65$ (Fig. \ref{fig:DeltaF}). A similar study on LJ systems showed similar trends, with free energy barriers vanishing at $\Theta = 0.7$ and increasing linearly from $\Theta = 0.7$ to $0.8$ \cite{Trudu2006}. However, as we will discuss in section \ref{sec:FES_interpretation}, $F(Q)$ barriers do not correspond to the CNT free energy barrier. $F(N)$ does not vanish, so interpret this vanishing free energy barrier as a signal of spinodal decomposition is incorrect. 

\subsection{Interpretation of the Free Energy Surface}
\label{sec:FES_interpretation}

We have outlined above how we use metaD to find the free energy surface $F(Q)$. However, we indicated in section \ref{sec:CNT} that the relevant quantity for describing classical nucleation is $F(N)$. In this section we explain why the easiest-to-access free energies are not the correct ones to use to calculate nucleation rates, but we can use Porion's recent analysis of cluster size distributions \cite{Porion2024} to obtain the correct free energy surface. 

To elucidate the issue, we consider three different, but related distributions and their associated free energies.
\begin{enumerate}
    \item $P(Q) \propto e^{-F(Q)/k_BT}$, the probability that the system's order parameter has the value $Q$.
    \item $P_\text{l}(N) \propto N_\text{tot}e^{-F_\text{l}(N)/k_BT}$, the probability that, in a system of $N_\text{tot}$ particles, the largest cluster contains $N$ particles. 
    \item $P_\text{a}(N) \propto N_\text{tot}e^{-F_\text{a}(N)/k_BT}$, the probability that a randomly chosen cluster is of size $N$\footnote{This definition is not precise. The ``probability'' is really something more like the average number of solid clusters of size $N$ when there are $N_\text{tot}$ total particles. This is also proportional (in the case of rare clusters) to the probability that any particle chosen at random is a member of a cluster of size $N$. When nearly all of the particles are in the liquid/monomer state, this further reduces to the probability given in the text.}.
\end{enumerate}
These first two are easily and commonly calculated\cite{tenWolde1996, Auer2001, Auer2004, Valeriani2005, Trudu2006, Tribello2017}, but the third is the proper distribution to use in CNT. In the case where N is large and there is only one cluster in the simulation volume, these distributions are all similar. This fact is sometimes used to justify using $F(Q)$ or $F_\text{l}(N)$ for nucleation rate calculations \cite{Kelton1991, Reiss1999}.

Unfortunately, these distributions differ when $N$ is small ($\sim 10$s of particles) and the system is near the bottom of the liquid free energy well \cite{Quiggley2009}. $F(Q)$ characterizes the system overall. The value of $F(Q)$ at the bottom of the liquid well is the free energy of the liquid state, including solid-like fluctuations around equilibrium. $F_\text{l}(N)$ characterizes a state where a liquid region may have multiple solid clusters, but the largest cluster has exactly $N$ particles. This depends strongly on the size of the simulation because large liquid regions are more likely to have many solid clusters, one of which may exceed $N$ constituent particles \cite{Porion2024}. $F_\text{a}(N)$ is the free energy of a particular cluster containing $N$ particles surrounded by a bulk liquid. The later is the correct free energy that corresponds to the CNT free energy in section \ref{sec:CNT}.

\begin{figure}
    \centering
    \includegraphics[width=1\linewidth]{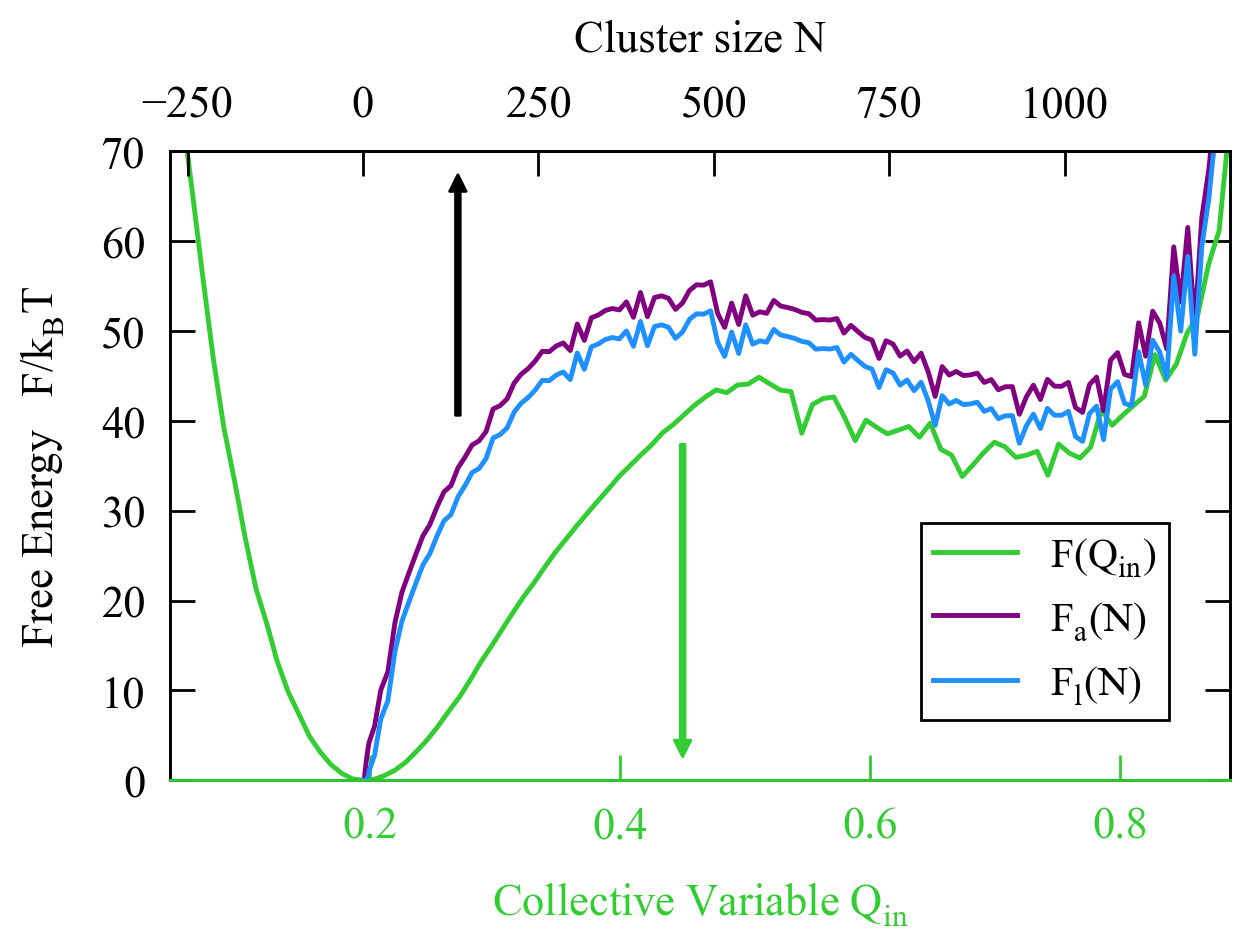}
    \caption{Changes in the $\kappa=0$, $\Theta=0.8$ FES due to our analysis procedure. The left well around $N=0$ and $Q_6=0.2$ is the liquid well. On the right there is a well corresponding to the solid which appears shallower than the liquid well due to our application of restraining wall potentials (see supplementary material). Each correction increases the free energy barrier height so the correct (purple) free energy barrier is $\sim 10k_BT$ higher than the initial metadynamics output (green).}
    \label{fig:FES}
\end{figure}

It is important that these distributions differ for small $N$ because the small-$N$ behavior of the distributions determines their normalization. Therefore, even when the large-$N$ behavior of the distributions match, the easily-calculated free energies $F(Q)$ and $F_\text{l}(N)$ do not yield the same same barrier heights as the correct free energy $F_\text{a}(N)$. Porion and Puibasset, in reference \onlinecite{Porion2024}, derived and validated relations between $F_\text{l}(N)$ and $F_\text{a}(N)$ that allow conversion from $F_\text{l}(N)$ to $F_\text{a}(N)$. Figure \ref{fig:FES} shows that these free energy barriers can differ by $\sim10k_BT$, and that $F_\text{a}(N)$ gives the highest free energy barrier, so using the correct free energy surface will predict a smaller nucleation rate than the commonly-used free energy surfaces. 

One final issue arises because we allow crystals to form in only a small fraction ($N_\text{in}/N_\text{tot}$) of the simulation volume. When we calculate the probability of forming a critical cluster, $P_\text{a}\left(N^*\right)$ from metadynamics, we have actually calculated the probability of forming a critical cluster in one particular region of the simulation volume. The corrected probability of forming a critical cluster \emph{anywhere} is greater by a factor of $N_\text{tot}/N_\text{in}$. This correction propagates from the probability distribution to the associated free energy \cite{Trudu2006} because $F_\text{a}(N)  = -k_BT\ln\big(P_\text{a}(N)\big)$.

\begin{align}
    P_\text{a}\left(N^*\right) &\rightarrow  \frac{N_\text{tot}}{N_\text{in}}  P_\text{a}\left(N^*\right) \\
    F_\text{a}\left(N^*\right) &\rightarrow  F_\text{a}\left(N^*\right)  -k_B T\ln{\left(\frac{N_\text{tot}}{N_\text{in}}\right)}
\end{align}
This correction occurs in step \ref{finite_size_correction} of the procedure in the next section. It causes a small change of roughly $1-5\%$ in our free energy barrier heights.

\subsection{Nucleation Barrier Calculation}
\label{sec:procedure}

In order to correctly incorporate observations from the previous few sections, our procedure for calculating the nucleation free energy barrier is as follows:
\begin{enumerate}
    \item Do metadynamics biasing the local CV $Q_\text{in}$ to calculate $F(Q_\text{in})$
    \item Use PLUMED functionality to calculate the largest cluster in the simulation at each timestep \cite{Tribello2017} 
    \item Use balanced exponential reweighting \cite{Schafer2020} to convert $F(Q_\text{in})\rightarrow F_\text{l}(N)$
    \item  Use equations 2 and 9 of [\onlinecite{Porion2024}] to convert $F_\text{l}(N) \rightarrow F_\text{a}(N)$
    \item Shift the FES to $\Delta F(N) = F_\text{a}(N)-F_\text{a}(0)$.
    \item  \label{finite_size_correction}Multiply $\Delta F(N)$ by $1+\frac{\ln{\left(N_\text{in}/N_\text{tot}\right)}}{\Delta F\left(N^*\right)}$. 
\end{enumerate}

Note that while this procedure improves the theoretical consistency between metadynamics results and the CNT framework, there is work to be done to ensure that post-processing procedure does not introduce any unexpected errors. For example, the criteria used to identify clusters are somewhat arbitrary, meaning that there is ambiguity in whether small clusters are really solid, or just ordered liquid regions. This ambiguity affects the normalization of $F_\text{a}(N)$ after the application of Porion's formulas. Eradicating these problems is difficult, so our later comparisons to benchmark simulations are vital to ensure the results of this analysis are trustworthy.

Specific parameters used for the metadynamics simulations are in the supplementary materials. Figure \ref{fig:FES_fits}, further discussed below, shows the FES calculated this way for $\kappa=0$.

\section{Nucleation Rates}
\label{sec:nuc_rates}

The CNT nucleation rate (eq. \eqref{eq:nucRate}) is the product of a kinetic prefactor and the dominant $\exp{\left\{-\Delta F\left(N^*\right)/k_BT\right\}}$ term. We calculate the nucleation rate in two ways. First, in the metaD method, we take the kinetic term from ref. [\onlinecite{Arnold2025}] and $\Delta F(N^*)$ from fits performed on the FES calculated in section \ref{sec:metaD}. Second, in order to validate the metaD method, we extract the full nucleation rate directly from low-temperature brute force simulations.

\subsection{Nucleation Rates from MetaD}
\label{sec:metaD_nucleation}

Assuming that $\Delta F(N)$ takes the form of Eq. \eqref{eq:freeEnergy}, there are two parameters that characterize the FES: the bulk solid-liquid free energy difference per particle $\Delta \mu$ and the interfacial energy coefficient $\gamma$. We take $\Delta \mu \approx \Delta f$, the Helmholtz free energy difference \footnote{This neglects volume differences between the solid and liquid, which is valid for astrophysical systems. In a system with important volume differences between the phases as constant pressure, the correct expression would be $\Delta \mu = \Delta f + P \Delta v$} per particle between the solid and liquid from refs. \onlinecite{Farouki1994, HamaguchiTriplePoint1997}. This leaves only the unknown surface term $\gamma$ to be determined via fits to our calculated free energy surfaces.

Rather than assuming surface energy is the same for clusters of any size, we use a Tolman length correction. As clusters become smaller, their surfaces become more curved. The curvature of a sphere is $1/R \propto N^{-1/3}$, so a first order expansion of the interfacial free energy in the curvature is 
\begin{equation}
\label{eq:Tolman}
    \gamma = \gamma_\infty \left(1-\frac{\delta}{N^{1/3}}\right),
\end{equation}
where $\gamma_\infty$ is the interfacial free energy coefficient for a planar liquid-solid interface and the Tolman parameter $\delta$ corrects the interfacial free energy of small clusters with high curvature. Using the tolman correction in Equation \eqref{eq:freeEnergy}, the CNT free energy surface becomes
\begin{equation}
    \label{eq:Tolman_freeEnergy}
    \Delta F(N) = \Delta \mu N  + \gamma_\infty \left(1-\frac{\delta}{N^{1/3}}\right) N^{2/3}
\end{equation}
with two unknown parameters $\gamma_\infty$ and $\delta$. This form of the free energy describes clusters as small as 10s of particles \cite{Chen2025} and the CNT assumptions underlying it become better for larger clusters \cite{Moroni2005}. 

\begin{figure}
    \centering
    \includegraphics[width=1\linewidth]{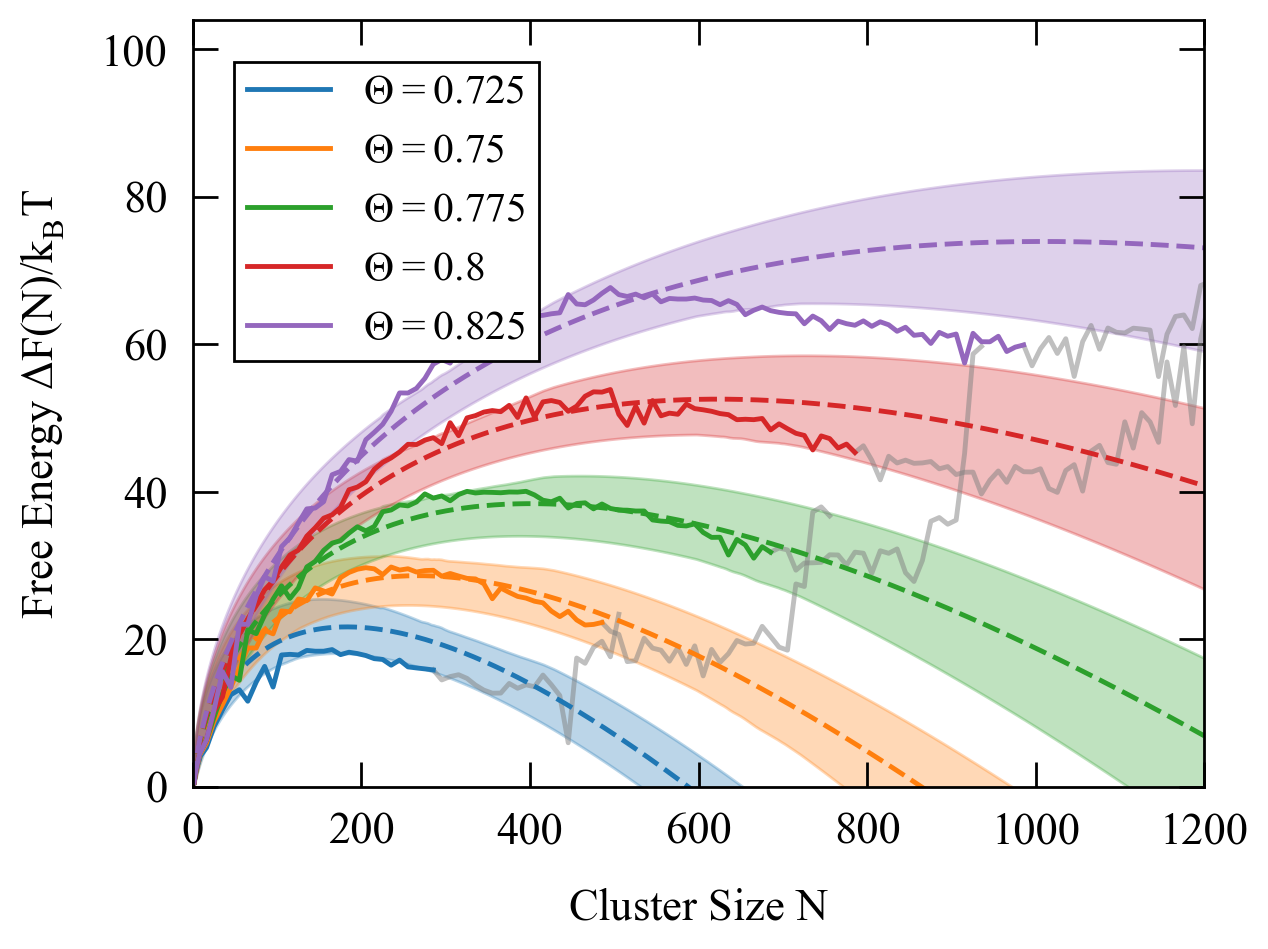}
    \caption{$\kappa=0$ free energy surfaces at different temperatures, represented by color. In blue, the lowest temperature has the smallest free energy barrier because because the liquid is less metastable at low temperatures. The gray portions were discarded for the fits because they are not physically relevant (see supplementary material). Dashed lines show fits to the metaD FES and the uncertainty bands are uncertainties due to the fitting procedure.}
    \label{fig:FES_fits}
\end{figure}

We get one value of  $\gamma_\infty$ and $\delta$ by simultaneously fitting Eq \ref{eq:Tolman_freeEnergy} to all free energy surfaces for each $\kappa$. The fitted parameters generated for all $\kappa$ are in table \ref{tab:fit_params} and the $\kappa=0$ fits to metaD data are shown in Fig. \ref{fig:FES_fits}.

Experimental determinations of the ratio between surface energy and enthalpy of fusion of liquid metals \cite{Turnbull1950}, combined with the enthalpy of fusion for the OCP\cite{Cooper2008} suggests $\gamma_\infty  \approx 1.7$. Interestingly, our values in Table \ref{tab:fit_params} are in reasonable agreement with the empirical estimate of surface energy.

We have found that the parameters fitted to a single FES are comparable to those calculated from simultaneous fits to all FES. Hence, a calculation at a single temperature for each $\kappa$ could be used with Equation \eqref{eq:Tolman_freeEnergy} to predict the FES at any other temperature. This method uses a small amount of simulation data to gain insight about about nucleation at different conditions. This efficiency makes our method appealing to search for trends in nucleation rate in large parameter spaces.

\begin{figure}
    \centering
    \includegraphics[width=1\linewidth]{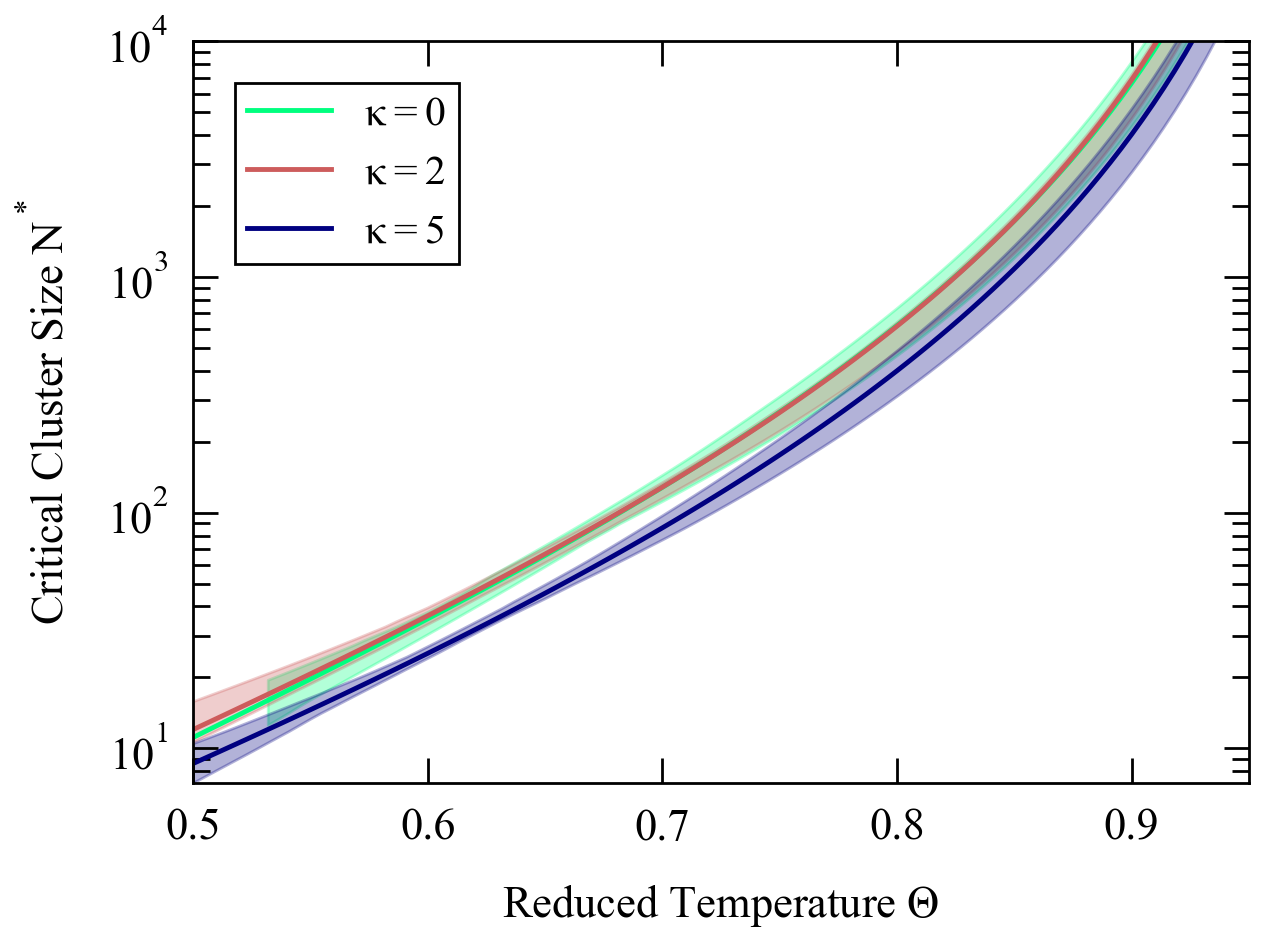}
    \caption{Critical cluster size obtained from the fits to metaD free energies. Colors correspond to the different screening lengths of the interatomic potential. The critical cluster size does not depend strongly on $\kappa$, and it becomes small (10s of particles) around $\Theta = 0.65$, the minimum temperature where we should expect a simple CNT+Tolman form to describe the free energy surface. $\kappa = 0$ and $2$ have similar critical cluster sizes, but $\kappa=5$ (blue) is consistently smaller.}
    \label{fig:N_star}
\end{figure}

Equation \eqref{eq:Tolman_freeEnergy} has a maximum at the critical cluster size $N^*$ (plotted in Fig. \ref{fig:N_star}). This maximum defines the free energy barrier height $\Delta F(N^*)$, which is plotted for comparison with other results in Fig. \ref{fig:bruteForce}b, which will be described later.  Equation \eqref{eq:nucRate} gives the nucleation rate based on these barriers.

\begin{table}[]
    \centering
    \begin{ruledtabular}
    \begin{tabular}{clcll}
 & \multicolumn{2}{c}{Large-N Fits}& \multicolumn{2}{c}{Small-N Fits}\\
         $\kappa$&   $\gamma_\infty$& $\delta$& $\gamma_\infty$&$\delta$\\ \hline
 0& 2.45& 0.51& 2.11&0.69\\
 2& 2.32& 0.44& 2.02&0.67\\
         5& 
     1.94& 0.01& 1.67&0.41\\ 
     \end{tabular}
     \end{ruledtabular}
    \caption{CNT+Tolman fitting parameters to describe the $F_a(N)$ free energy barrier. The large-N fits describe the formation of large clusters and nucleation behavior of the system. For all screening lengths, $\gamma_\infty$ is smaller and $\delta$ is larger when the fits are performed on only the small-N part of the free energy surface. The discrepancy between the two sets of parameters indicates that the free energy barrier required to form a very small cluster is less than CNT would predict, so the overall fits do not apply at low temperatures where only small clusters are relevant.}
    \label{tab:fit_params}
\end{table}

\subsection{Nucleation Rates from Brute Force Simulations}
\label{sec:brute_force_nucleation}

Brute force simulations, wherein molecular dynamics simulations of supercooled liquids evolve freely until they solidify, are the most accurate way to calculate nucleation rates. However, these simulations are only feasible under deep supercooling when nucleation proceeds quickly. Still, they provide gold-standard nucleation rates at low temperatures to which we will compare metaD predictions.

\begin{figure}
    \centering
    \includegraphics[width=1\linewidth]{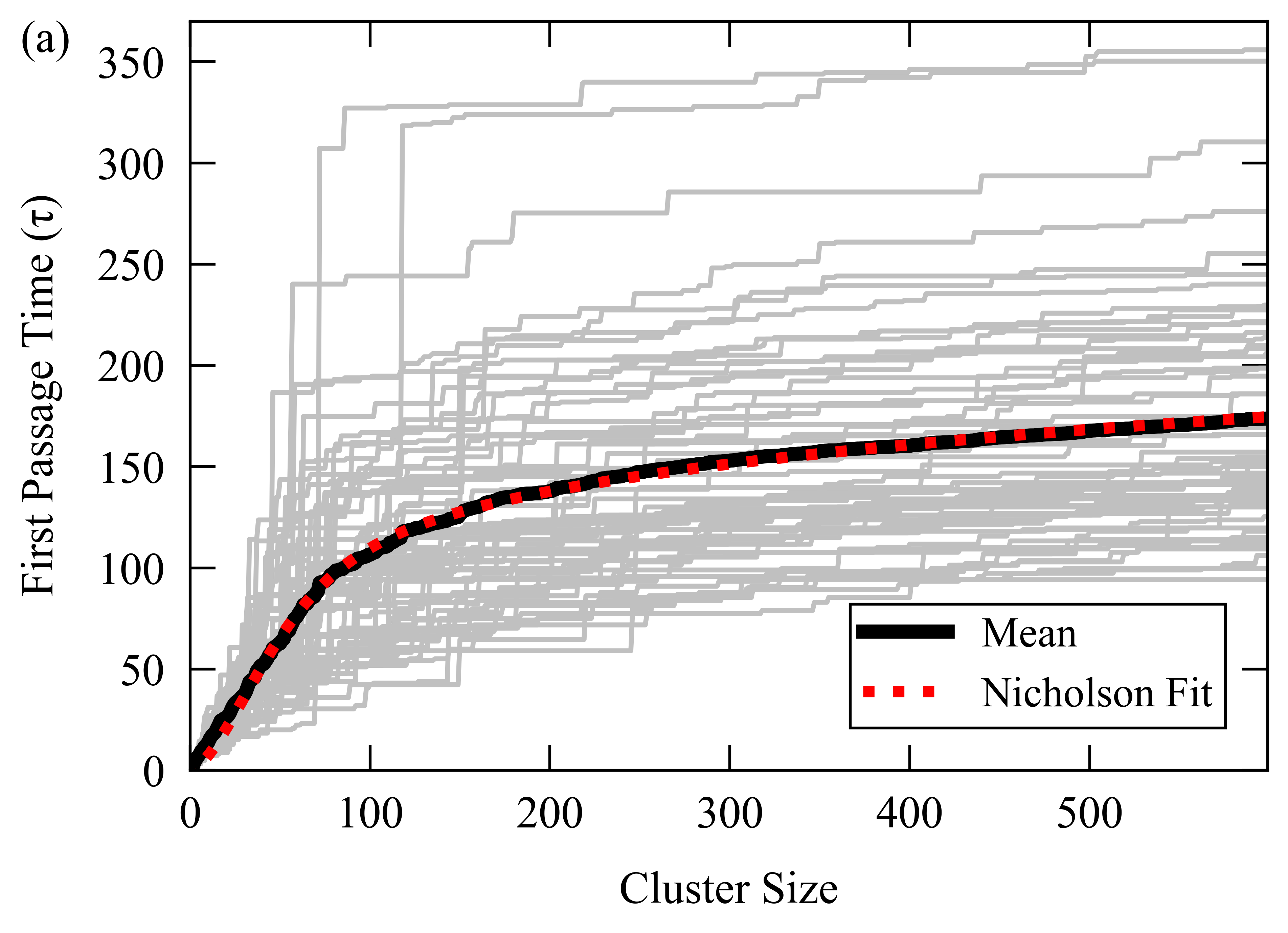}
    \includegraphics[width=1\linewidth]{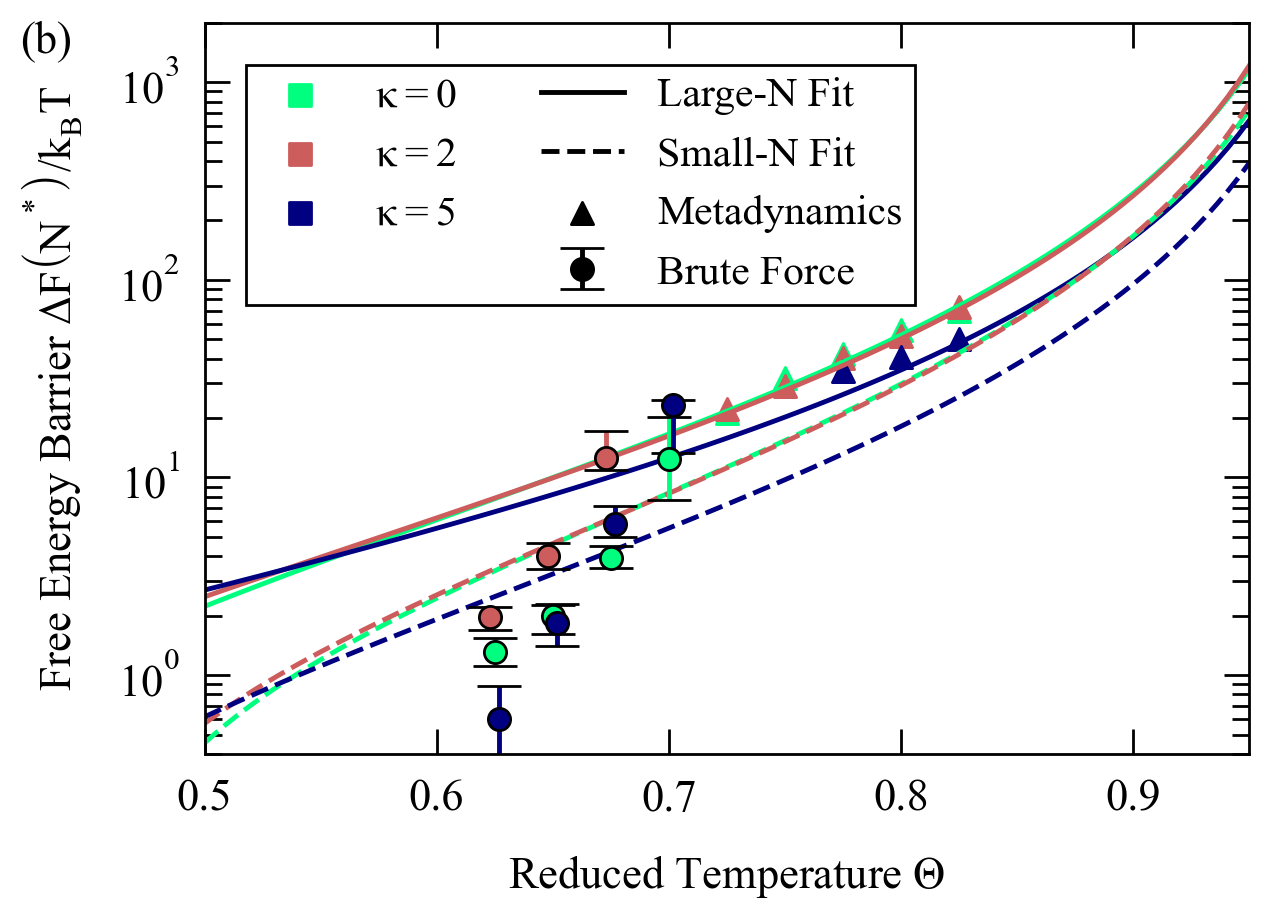}
    \caption{(a) illustrative first passage time data, including fits to Nicholson's $i=2$ cummulant expansion \cite{Nicholson2016}. (b) free energy barriers from brute force simulations (circles) and metaD (triangles) compared to fits performed on all cluster sizes (solid lines) and only small clusters (dashed lines). Colors correspond to different screening lengths. For all $\kappa$, the fits performed on all cluster sizes describe the metadynamics results and the free energy barriers from the highest-temperature brute force simulations. These fits should also describe nucleation at higher temperatures, closer to the melt temperature. The lower temperature brute force simulations require modified CNT+Tolman parameters from fits to small clusters.}
    \label{fig:bruteForce}
\end{figure}

At several temperatures for each $\kappa$, we run many simulations of independently initialized supercooled liquids until they freeze naturally. We calculate the first passage time $\tau(N)$, the time at which a solid cluster of size $N$ first appears, for each simulation. The gray curves in Fig. \ref{fig:bruteForce}a display an ensemble of first passage times from $50$ independent simulations.  The mean of these first passage times (the MFPT, black curve in Fig. \ref{fig:bruteForce}a) contains information about the nucleation rate and crystal growth rate at the $\kappa, \Theta$ condition where the simulations were performed. 

To extract the nucleation rate from the MFPT, one must model the FPT distribution and fit that modeled distribution to the simulations. Typically, this is done under the assumption that crystal growth is rapid relative to the rate of nucleation \cite{wedekind2007}. This separation of timescales applies in this system only for $\Theta > 0.65$, so we use a more flexible model. Nicholson [\onlinecite{Nicholson2016}] calculates a cummulant expansion of the FPT distribution as a function of the free energy barrier, the critical cluster size, and a kinetic factor. We use the second-order cummulant expansion of the first passage time distribution (see appendix B of [\onlinecite{Nicholson2016}]). This expression fits our MFPT well (Fig. \ref{fig:bruteForce}a, dashed red curve), providing estimates of nucleation rates and free energy barriers. We plot these estimated free energy barriers at low temperatures in Fig. \ref{fig:bruteForce}b.

We compare the brute force free energy barriers to different fits performed on metaD. In section \ref{sec:metaD_nucleation} we find the values of $\gamma_\infty$ and $\delta$ that best describe the overall metaD free energy surfaces. These fitted values give the barrier heights shown as solid lines in Fig. \ref{fig:bruteForce}b, which agree well with the individual metaD barriers (triangles), but overestimate the barriers from brute force simulations (circles). 

To understand this discrepancy, we do additional fits to the metaD data, considering only small clusters ($N\leq 30$). These fits (dashed curves) on the small-N portion of the \emph{high temperature} data intersect the brute force data at \emph{low} temperature. This agreement is significant because it suggests that the small-N parts of our free energy surfaces contain information about the small clusters relevant to low temperature nucleation. However, the very smallest clusters are highly irregular or ill-defined (Fig. \ref{fig:CNT_FES}a), so CNT with a Tolman correction can only describe free energy barriers below $\Theta=0.7$ qualitatively. An expansion to higher orders in $N^{-1/3}$ in Equation \eqref{eq:Tolman} may more accurately describe very small clusters \cite{Gispen2024.2}.

Other work on nucleation free energy surfaces shows that if a set of CNT+Tolman parameters describe a nucleation free energy surface for moderate cluster sizes, they also describe large clusters. Since our CNT+Tolman fits describe all data $\Theta \geq 0.7$ (including the highest temperature brute force points) we conclude that our fits are extensible to our applications of interest in the higher-temperature regime (i.e. larger critical clusters).


\subsection{Nucleation Rate Comparison}

\begin{figure}
    \centering
    \includegraphics[width=1\linewidth]{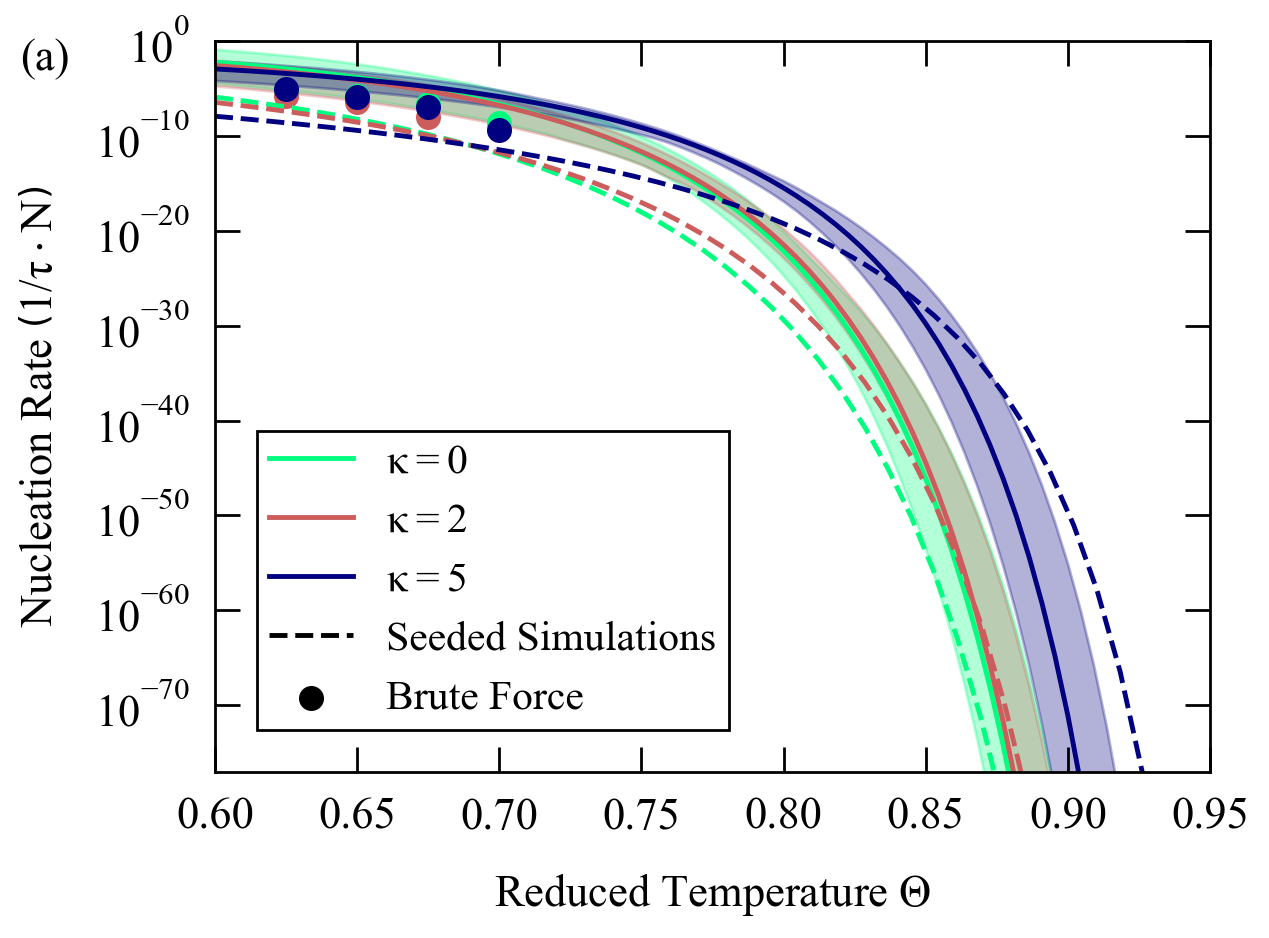}
    \includegraphics[width=1\linewidth]{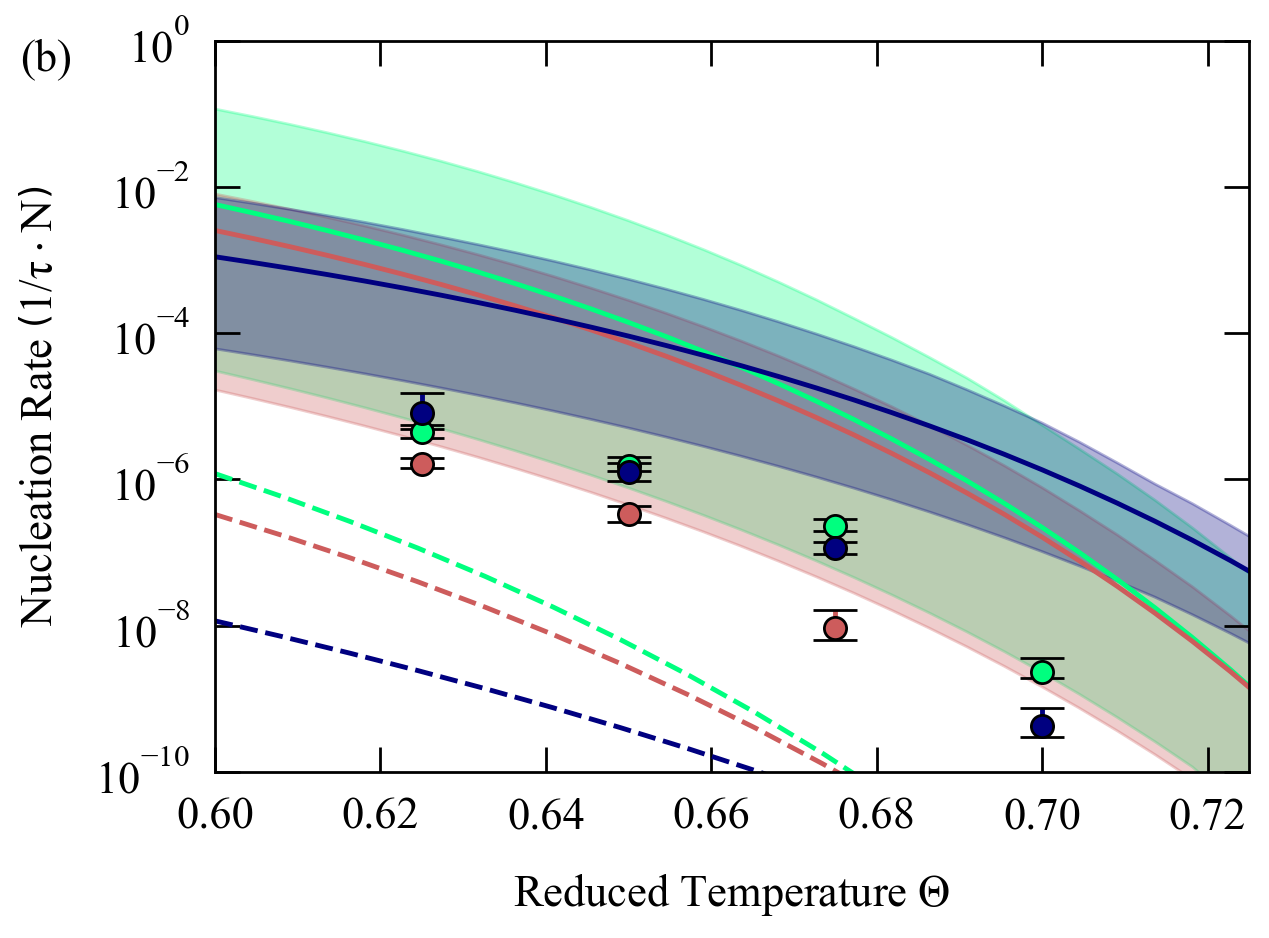}
    \caption{Comparison of nucleation rates calculated using different methods, including the brute force (points) and metadynamics procedures we use in this paper (solid lines), and seeded simulations from [\onlinecite{Arnold2025}] (dashed lines). (a) The nucleation rate is exponentially suppressed as $\Theta \rightarrow 1$. At high temperatures the $\kappa =5$ system (blue) with the shortest interaction range nucleates faster than the more strongly interacting systems. This corresponds to the smaller free energy barrier shown in Figure \ref{fig:bruteForce}. (b) shows the comparison between the gold-standard brute force simulations and the other methods.}
    \label{fig:nuc_rates}
\end{figure}

From Equation \eqref{eq:nucRate} and the free energy fits described in section \ref{sec:metaD_nucleation}, we get nucleation rates for $\kappa =0, 2,$ and $5$ at a range of temperatures. Fig. \ref{fig:nuc_rates} plots these nucleation rates with uncertainty bands. The top panel shows trends in nucleation rates from rapid nucleation on MD timescales for $\Theta<0.7$ to extremely slow nucleation as $\Theta$ approaches $1$. Qualitatively, the metaD fits are similar to the nucleation rates obtained from fits to seeded simulations \cite{Arnold2025}. We see that $\kappa=0$ and $2$ have similar nucleation rates, while $\kappa=5$ nucleates much faster where it has a lower free energy barrier at high temperatures. All $\kappa$ have similar nucleation rates, when scaled by a thermal timescale $\tau$, at low temperatures where barriers are small and kinetics dominate the nucleation rates.

The lower panel of Fig. \ref{fig:nuc_rates} shows the agreement between nucleation rates calculated by metaD and brute force calculations. The brute force nucleation rates are systematically below the metaD predictions, but Fig. \ref{fig:bruteForce}b shows that the $\Theta=0.7$ brute force simulation free energy barriers agree well with the metaD predictions. Since the nucleation rate is a product of the free energy barrier and a kinetic term, the discrepancy between brute force and metaD nucleation rates must originate in the kinetic term. The kinetic term becomes irrelevant at high temperatures, so we expect that our results are even more accurate at larger $\Theta$.

\section{Conclusion}

We have presented a metaD procedure to calculate the free energy barrier for formation of a crystal from a supercooled liquid. We used a local collective variable $Q_\text{in}$ to efficiently sample the classical nucleation pathway. This allowed us to run our simulations at a higher temperature than would be possible with a global collective variable. As is standard, we reweight $F(Q_\text{in})$ to estimate $F_\text{l}(N)$, the free energy associated with the simulation's largest cluster being of size $N$. We then showed that this free energy must be further corrected to $F_\text{a}(N)$, a free energy associated with the probability that any cluster is of size $N$,  in order to be consistent with classical nucleation theory. This final correction is novel and increases nucleation free energy barrier heights, leading to slower predicted nucleation rates. Finally, we fit the resulting CNT-consistent free energy surfaces to a CNT+Tolman form, providing a fit that predicts nucleation rates at temperatures beyond those where we did our simulations. 

These nucleation rates are reliable; the CNT+Tolman form of our fit explains the free energy barriers calculated by benchmark brute force calculations at low temperatures, and the fits become more accurate at higher temperatures \cite{Moroni2005, Chen2025}. The accuracy of our method could be assessed with additional comparisons to other methods, such as path sampling \cite{Sosso2016, Finney2023} or infrequent metaD \cite{Tiwary2013}. However, these methods do not provide free energy surfaces that can be fit and extrapolated as easily as ours, which limits their predictive scope.

The steep computational cost of calculating $Q_6$ for all particles limits this method to simulations of several thousand particles. A collective variable that can be calculated faster, possibly based on machine learning, or an on-the-fly CV optimization scheme could allow application of our method to more systems, including those with complex crystallization pathways or complex structures with many polymorphs \cite{Bussi2020, Desgranges2025}. Alternate umbrella sampling approaches optimized to measure CNT-like free energy surfaces could also speed up calculations \cite{Gispen2024.2}.

When this method is applied to the YOCP, the resulting nucleation rates show trends consistent with our previous estimates from seeded simulations, but the improved nucleation rates are slightly higher at low temperatures. They confirm our prior observations that the weakly screened YOCP with long-range interactions must be supercooled around 15\% before nucleation rates are high enough for solidification to begin \cite{Arnold2025}. Systems with stronger screening (shorter interaction ranges) have faster nucleation rates and require slightly less supercooling. These results are very informative for slowly cooling systems where the critical temperature (the temperature at which the nucleation rate becomes fast) is more important than the nucleation rate itself. The horizontally narrow uncertainty bands in figure \ref{fig:nuc_rates}a show that, despite uncertainty on fitting parameters, the critical temperature is well-constrained.

With this method in place to predict nucleation rates of the YOCP, nucleation rates of more complex systems are within reach. This procedure to be applied to other systems by choosing a different CV in place of $Q_6$. The benefits of local CVs, the $F(Q_\text{in})\rightarrow F_\text{l}(N)\rightarrow F_\text{a}(N)$ corrections, and our fitting procedure will benefits studies of these other systems too by providing free energy surfaces that are consistent with classical nucleation theory, and therefore suitable for predicting nucleation rates for any system.

Multi-component systems, which are only beginning to be understood \cite{Iida2023, Yamamura2025, Aslanov2025}, are one possible target. We plan to pursue Coulomb plasma mixtures with astrophysical relevance. Crystallization of mixtures in old white dwarf stars is vital to to understanding astrophysical observations \cite{caplan2020, Bedard2024}. However, previous work neglects microscopic details of this crystallization. This makes white dwarf astrophysics a compelling, if surprising, target for applying the recent advances in understanding nucleation. 

\section*{Supplementary Material}
We provide in \url{https://doi.org/10.6084/m9.figshare.31164304} LAMMPS scripts, PLUMED scripts, and PLUMED outputs for representative metaD simulations used to generate $F_\text{l}(N)$. Data files containing $F_\text{a}(N)$ for all $\kappa = 0, 2,$ and $5$ are in the repository as well.

The supplementary materials file contains additional information about the exact parameters used to run metadynamics, including the use of restraining wall potentials, and the estimation of uncertainty associated with the fitting procedure.

\begin{acknowledgments}
This material is based upon work supported by the Department of Energy (National Nuclear Security Administration) University of Rochester “National Inertial Confinement Fusion Program” under Award No. DE-NA0004144. The work of J.D. and D.S. was performed under the auspices of the U.S. Department of Energy under Contract No. 89233218CNA000001, and was supported by the Laboratory Directed Research and Development program of Los Alamos National Laboratory.

We thank J. X. D'Souza for his thoughful feedback on this manuscript.
\end{acknowledgments}

\section*{Author Declarations}
\subsection*{Conflicts of Interest}
The authors have no conflicts to disclose
\subsection*{Author Contributions}
\textbf{Brennan Arnold:} conceptualization (equal), formal analysis (lead), visualization (lead), writing- orignal draft preparation (lead), writing- review and editing (equal) 
\textbf{Jerome Daligault:} conceptualization (equal), writing- review and editing (equal) 
\textbf{Didier Saumon:} conceptualization (equal), writing- review and editing (equal) 
\textbf{Suxing Hu:} conceptualization (equal), writing- review and editing (equal) 
\subsection*{Data Availability statement}
The data that support the findings of this study are openly available at \url{https://doi.org/10.6084/m9.figshare.31164304} 

\bibliography{YOCPmetaD}

@PREAMBLE{
 "\providecommand{\noopsort}[1]{}" 
 # "\providecommand{\singleletter}[1]{#1}%" 
}

@ARTICLE{Arnold2025,
    title = {Crystal nucleation rates in one-component {Y}ukawa systems},
  author = {Arnold, B. and Daligault, J. and Saumon, D. and B\'edard, Antoine and Hu, S. X.},
  journal = {Phys. Rev. E},
  volume = {111},
  issue = {2},
  pages = {025206},
  numpages = {15},
  year = {2025},
  month = {Feb},
  publisher = {American Physical Society},
  doi = {10.1103/PhysRevE.111.025206},
  url = {https://link.aps.org/doi/10.1103/PhysRevE.111.025206}
}

@Article{LAMMPS,
    author = "A. P. Thompson and H. M. Aktulga and R. Berger and D. S. Bolintineanu and W. M. Brown and P. S. Crozier and P. J. in 't Veld and A. Kohlmeyer and S. G. Moore and T. D. Nguyen and R. Shan and M. J. Stevens and J. Tranchida and C. Trott and S. J. Plimpton",
    title = "{LAMMPS} - a flexible simulation tool for particle-based materials modeling at the atomic, meso, and continuum scales",
    journal = "Comput. Phys. Commun.",
    volume =  "271",
    pages =   "108171",
    year =    "2022",
    doi = "10.1016/j.cpc.2021.108171"
}

@ARTICLE{PLUMED,
    author = "Gareth A. Tribello AND Massimiliano Bonomi AND Davide Branduardi AND Carlo Camilloni AND Giovanni Bussi", 
    title = "Plumed 2: New feathers for an old bird", 
    journal = "Comput. Phys. Commun.",
    volume = "185", 
    pages = "604–613", 
    year = "2014"
}

@ARTICLE{Aslanov2025,
    author = "Leonid A. Aslanov AND Igor K. Kudryavtsev AND Sergey F. Dunaev ", 
    title = "Current state of the problem of crystal nucleation", 
    journal = "Struct Chem",
    volume = "36", 
    pages = "1983–1992", 
    year = "2025"
}

@ARTICLE{BC8,
    author = "Kien Nguyen-Cong AND Jonathan T. Willman AND Joseph M. Gonzalez AND Ashley S. Williams AND Anatoly B. Belonoshko AND Stan G. Moore AND Aidan P. Thompson AND Mitchell A. Wood AND Jon H. Eggert AND Marius Millot AND Luis A. Zepeda-Ruiz AND Ivan I. Oleynik", 
    title = "Extreme Metastability of Diamond and its Transformation to the BC8 Post-Diamond Phase of Carbon", 
    journal = "J. Phys. Chem. Lett.",
    volume = "15", 
    pages = "1152–1160", 
    year = "2024"
}

@ARTICLE{Kraus2017,
    author = "D. Kraus AND J. Vorberger AND A. Pak AND N. J. Hartley AND L. B. Fletcher AND S. Frydrych AND E. Galtier AND E. J. Gamboa AND D. O. Gericke AND S. H. Glenzer AND E. Granados AND M. J. MacDonald AND A. J. MacKinnon AND E. E. McBride AND I. Nam AND P. Neumayer AND M. Roth AND A. M. Saunders AND A. K. Schuster AND P. Sun AND T. van Driel AND T. Döppner AND R. W. Falcone", 
    title = "Formation of diamonds in laser-compressed hydrocarbons at planetary interior conditions", 
    journal = "Nat Astron",
    volume = "1", 
    pages = "606-611", 
    year = "2017"
}

@ARTICLE{Shi2023,
    author = "Jiuyang Shi AND Zhixing Liang AND Junjie Wang AND Shuning Pan AND Chi Ding AND Yong Wang AND Hui-Tian Wang AND Dingyu Xing AND Jian Sun", 
    title = "Double-Shock Compression Pathways from Diamond to BC8 Carbon", 
    journal = "Phys. Rev. Lett.",
    volume = "131", 
    pages = "146101", 
    year = "2023"
}

@ARTICLE{Gao2025,
    author = "Chen Gao AND Kai-Ming Ho AND Renata M. Wentzcovitch AND Yang Sun", 
    title = "Understanding the two-step nucleation of iron at Earth's inner core conditions: A comparative molecular dynamics study", 
    journal = "Phys. Rev. B",
    volume = "111", 
    pages = "134104", 
    year = "2025"
}

@ARTICLE{Huguet2018,
    author = "Ludovic Huguet AND James A. Van Orman AND Steven A. Hauck II AND Matthew A. Willard", 
    title = "Earth's inner core nucleation paradox", 
    journal = "Earth and Planetary Science Letters",
    volume = "487", 
    pages = "9-20", 
    year = "2018"
}

@ARTICLE{wilson2023,
    author = "Alfred J. Wilson AND Dario Alfè AND Andrew M. Walker AND Christopher J. Davies", 
    title = "Can homogeneous nucleation resolve the inner core nucleation paradox?", 
    journal = "Earth and Planetary Science Letters",
    volume = "614", 
    pages = "118176", 
    year = "2023"
}

@ARTICLE{Tremblay2019,
    author = "Pier-Emmanuel Tremblay AND Gilles Fontaine AND Nicola Pietro Gentile Fusillo AND Bart H. Dunlap AND Boris T. Gänsicke AND Mark A. Hollands AND J. J. Hermes AND Thomas R. Marsh AND Elena Cukanovaite AND Tim Cunningham ", 
    title = "Core crystallization and pile-up in the cooling sequence of evolving white dwarfs", 
    journal = "Nature",
    volume = "565", 
    pages = "202–205", 
    year = "2019"
}

@ARTICLE{Qbranch,
    author = "Sihao Cheng AND Jeffrey D. Cummings AND Brice Ménard", 
    title = "A Cooling Anomaly of High-mass White Dwarfs", 
    journal = "ApJ",
    volume = "886", 
    pages = "100", 
    year = "2019"
}

@ARTICLE{blouinNeDist2021,
    author = "Simon Blouin AND Jérôme Daligault AND Didier Saumon", 
    title = "22Ne Phase Separation as a Solution to the Ultramassive White Dwarf Cooling Anomaly", 
    journal = "ApJL",
    volume = "911", 
    pages = "L5", 
    year = "2021"
}

@ARTICLE{Bedard2024,
   author       = "Antoine Bédard AND Simon Blouin AND S. Cheng",
   title        = "Buoyant crystals halt the cooling of white dwarf stars",
   journal      = "Nature",
   volume       = "627", 
   pages        = "286-288",
   year         = "2024",
}

@ARTICLE{blow2021,
    author = "Katarina E. Blow AND David Quigley AND Gabriele C. Sosso", 
    title = "The seven deadly sins: When computing crystal nucleation rates, the devil is in the details", 
    journal = "J. Chem. Phys.",
    volume = "155", 
    pages = "040901", 
    year = "2021"
}

@ARTICLE{Sosso2016,
    author = "Gabriele C. Sosso AND Ji Chen AND Stephen J. Cox AND Martin Fitzner AND Philipp Pedevilla AND Andrea Zen AND Angelos Michaelides", 
    title = "Crystal Nucleation in Liquids: Open Questions and Future Challenges in Molecular Dynamics Simulations", 
    journal = "Chem. Rev.",
    volume = "116", 
    pages = "7078–7116", 
    year = "2016"
}

@ARTICLE{Finney2023,
    author = "Aaron R. Finney AND Matteo Salvalaglio", 
    title = "Molecular simulation approaches to study crystal nucleation from solutions: Theoretical considerations and computational challenges", 
    journal = "WIREs Computational Molecular Science",
    volume = "14", 
    pages = "e1697", 
    year = "2023"
}

@ARTICLE{Lutsko2019,
    author = {James F. Lutsko },
    title = {How crystals form: A theory of nucleation pathways},
    journal = {Science Advances},
    volume = {5},
    number = {4},
    pages = {eaav7399},
    year = {2019},
    doi = {10.1126/sciadv.aav7399}
}

@ARTICLE{Desgranges2025,
    author = "Caroline Desgranges AND Jerome Delhommelle.", 
    title = "Deciphering the complexities of crystalline state(s) with molecular simulations", 
    journal = "Commun Chem",
    volume = "8", 
    pages = "281", 
    year = "2025"
}

@ARTICLE{magsumov2024,
    author = "Timur Magsumov AND Ilya Ibraev AND Igor Sedov", 
    title = "Probing the Conformational Ensemble of the Amyloid Beta 16–22 Fragment with Parallel-Bias Metadynamics", 
    journal = "J. Phys. Chem. B",
    volume = "128",
    issue = "50",
    pages = "12333–12347", 
    year = "2024"
}

@ARTICLE{Santos2022,
    author = "Pedro A. Santos-Florez AND Howard Yanxon AND Byungkyun Kang AND Yansun Yao AND Qiang Zhu", 
    title = "Size-Dependent Nucleation in Crystal Phase Transition from Machine Learning Metadynamics", 
    journal = "Phys. Rev. Lett.",
    volume = "129", 
    pages = "185701", 
    year = "2022"
}

@ARTICLE{martonak2003,
    author = "R. Martoňák AND A. Laio AND M. Parrinello", 
    title = "Predicting Crystal Structures: The Parrinello-Rahman Method Revisited", 
    journal = "Phys. Rev. Lett",
    volume = "90", 
    pages = "075503", 
    year = "2003"
}

@ARTICLE{Badin2021,
    author = "Matej Badin AND Roman Martoňák", 
    title = "Nucleating a Different Coordination in a Crystal under Pressure: A Study of the B1-B2 Transition in NaCl by Metadynamics", 
    journal = "Phys. Rev. Lett.",
    volume = "127", 
    pages = "105701", 
    year = "2021"
}

@article{Giberti2015,
author = "Giberti, Federico and Salvalaglio, Matteo and Parrinello, Michele",
title = "{Metadynamics studies of crystal nucleation}",
journal = "IUCrJ",
year = "2015",
volume = "2",
number = "2",
pages = "256--266",
month = "Mar",
doi = {10.1107/S2052252514027626},
url = {https://doi.org/10.1107/S2052252514027626},
}

@ARTICLE{Tribello2017,
    author = "Gareth A. Tribello AND Federico Giberti AND Gabriele C. Sosso AND Matteo Salvalaglio AND Michele Parrinello", 
    title = "Analyzing and Driving Cluster Formation in Atomistic Simulations", 
    journal = "J. Chem. Theory Comput.",
    volume = "13", 
    issue = "3",
    pages = "1317–1327", 
    year = "2017"
}

@ARTICLE{Bussi2020,
    author = "Giovanni Bussi AND Alessandro Laio", 
    title = "Using metadynamics to explore complex free-energy landscapes", 
    journal = "Nature Reviews Physics",
    volume = "2", 
    pages = "200–212", 
    year = "2020"
}

@ARTICLE{Schafer2020,
    author = "Timo M. Schäfer AND Giovanni Settanni", 
    title = "Data Reweighting in Metadynamics Simulations", 
    journal = "J. Chem. Theory Comput.",
    volume = "16", 
    pages = "2042–2052", 
    year = "2020"
}

@ARTICLE{Tiwary2013,
    author = "Pratyush Tiwary AND Michele Parrinello", 
    title = "From Metadynamics to Dynamics", 
    journal = "Phys. Rev. Lett.",
    volume = "111", 
    pages = "230602", 
    year = "2013"
}

@ARTICLE{Trudu2006,
    author = "Federica Trudu AND Davide Donadio AND Michele Parrinello", 
    title = "Freezing of a Lennard-Jones Fluid: From Nucleation to Spinodal Regime", 
    journal = "Phys. Rev. Lett.",
    volume = "97", 
    pages = "105701", 
    year = "2006"
}

@ARTICLE{Branduardi2012,
    author = "Davide Branduardi AND Giovanni Bussi AND Michele Parrinello", 
    title = "Metadynamics with Adaptive Gaussians", 
    journal = "J. Chem. Theory Comput.",
    volume = "8", 
    pages = "2247–2254", 
    year = "2012"
}

@ARTICLE{Bonomi2009,
    author = "M. Bonomi AND A. Barducci AND M. Parrinello", 
    title = "Reconstructing the equilibrium Boltzmann distribution from well-tempered metadynamics", 
    journal = "J. Comput. Chem.",
    volume = "30", 
    pages = "1615-1621", 
    year = "2009"
}

@ARTICLE{Barducci2008,
    author = "Alessandro Barducci AND Giovanni Bussi AND Michele Parrinello", 
    title = "Well-Tempered Metadynamics: A Smoothly Converging and Tunable Free-Energy Method", 
    journal = "Phys. Rev. Lett.",
    volume = "100", 
    pages = "020603", 
    year = "2008"
}

@ARTICLE{Barducci2011,
    author = "Alessandro Barducci, Massimiliano Bonomi, Michele Parrinello", 
    title = "Metadynamics", 
    journal = "WIREs Comput Mol Sci",
    volume = "1", 
    pages = "826-843", 
    year = "2011"
}

@ARTICLE{Quiggley2009,
    author = "D. Quigley AND P.M. Rodger", 
    title = "A metadynamics-based approach to sampling crystallisation events", 
    journal = "Molecular Simulation",
    volume = "35", 
    pages = "613-623", 
    year = "2009"
}

@ARTICLE{Martonak2025,
    title={From diamond to BC8 to simple cubic and back: kinetic pathways to post-diamond carbon phases from metadynamics}, 
      author={Roman Martoňák and Sergey Galitskiy and Azat Tipeev and Joseph M. Gonzalez and Ivan I. Oleynik},
      year={2025},
      eprint={2509.00423},
      archivePrefix={arXiv},
      primaryClass={cond-mat.mtrl-sci},
      url={https://arxiv.org/abs/2509.00423}, 
}

@ARTICLE{Porion2024,
    author = "Patrice Porion AND Joël Puibasset", 
    title = "A statistical analysis of the first stages of freezing and melting of Lennard-Jones particles: Number and size distributions of transient nuclei", 
    journal = "J. Chem. Phys.",
    volume = "161", 
    pages = "074501", 
    year = "2024"
}

@ARTICLE{Bulutoglu2023,
    author = "Pelin S. Bulutoglu AND Akshat S. Zalte AND Nandkishor K. Nere AND Doraiswami Ramkrishna AND David S. Corti", 
    title = "A comprehensive modeling approach for polymorph selection in Lennard-Jones crystallization", 
    journal = "J. Chem. Phys.",
    volume = "158", 
    pages = "134505", 
    year = "2023"
}

@ARTICLE{Cheng2015,
    author = "Bingqing Cheng AND Gareth A. Tribello AND Michele Ceriotti", 
    title = "Solid-liquid interfacial free energy out of equilibrium", 
    journal = "Phys. Rev. B",
    volume = "92", 
    pages = "180102(R)", 
    year = "2015"
}

@ARTICLE{Cooper2008,
    author = "Randall L. Cooper and Lars Bildsten", 
    title = "Classical nucleation theory of the one-component plasma", 
    journal = "Phys. Rev. E",
    volume = "77", 
    pages = "056405", 
    year = "2008"
}

@ARTICLE{Reiss1999,
    author = "Howard Reiss; Richard K. Bowles", 
    title = "Some fundamental statistical mechanical relations concerning physical clusters of interest to nucleation theory", 
    journal = "J. Chem. Phys.",
    volume = "111", 
    pages = "7501–7504", 
    year = "1999"
}

@ARTICLE{Gispen2024.2,
    author = "Willem Gispen AND Jorge R. Espinosa AND Eduardo Sanz AND Carlos Vega AND Marjolein Dijkstra", 
    title = "Variational umbrella seeding for calculating nucleation barriers", 
    journal = "J. Chem. Phys.",
    volume = "160", 
    pages = "174501", 
    year = "2024"
}

@ARTICLE{Fu2022,
    author = "Haoyang Fu AND Xing Gao AND Xin Zhang AND Lan Ling", 
    title = "Recent Advances in Nonclassical Crystallization: Fundamentals,
Applications, and Challenges", 
    journal = "Cryst. Growth Des.",
    volume = "22", 
    pages = "1476-1499", 
    year = "2022"
}

@ARTICLE{Valeriani2005,
    author = "C. Valeriani AND E. Sanz AND D. Frenkel", 
    title = "Rate of homogeneous crystal nucleation in molten NaCl", 
    journal = "J. Chem. Phys.",
    volume = "122", 
    pages = "194501", 
    year = "2005"
}

@ARTICLE{Yamamura2025,
    author = "Takumu Yamamura AND Ryuhei Sato AND Yasushi AND Shibuta", 
    title = "Metadynamics for solid-liquid coexistence system of binary alloys", 
    journal = "Acta Materialia",
    volume = "297", 
    pages = "121348", 
    year = "2025"
}

@ARTICLE{Turnbull1950,
    author = "D. Turnbull", 
    title = "Formation of Crystal Nuclei in Liquid Metals", 
    journal = "J. Appl. Phys.",
    volume = "21", 
    pages = "1022–1028", 
    year = "1950"
}

@ARTICLE{Ghoufi2025,
    author = "Aziz Ghoufi", 
    title = "Atomistic computing of the solid–fluid surface free energy and tension", 
    journal = "Nature Reviews Physics",
    volume = "7", 
    pages = "473–486", 
    year = "2025"
}

@ARTICLE{Kelton1991,
    author = "K F Kelton", 
    title = "Crystal Nucleation in Liquids and Glasses", 
    journal = "Solid State Physics",
    volume = "45", 
    pages = "75-177", 
    year = "1991"
}

@ARTICLE{deJager2024,
    author = "Marjolein de Jager AND Carlos Vega AND Pablo Montero de Hijes AND Frank Smallenburg AND Laura Filion", 
    title = "Statistical mechanics of crystal nuclei of hard spheres", 
    journal = "J. Chem. Phys.",
    volume = "161", 
    pages = "184501", 
    year = "2024"
}

@ARTICLE{Arjun2023,
    author = "A. Arjun AND Peter G. Bolhuis", 
    title = "Homogeneous nucleation of crystalline methane hydrate in molecular dynamics transition paths sampled under realistic conditions", 
    journal = "J. Chem. Phys.",
    volume = "158", 
    pages = "044504", 
    year = "2023"
}

@ARTICLE{Lin2024,
    author = "Min Lin AND Zhewen Xiong AND Haishan Cao", 
    title = "Bridging classical nucleation theory and molecular dynamics simulation for homogeneous ice nucleation", 
    journal = "J. Chem. Phys.",
    volume = "161", 
    pages = "084504", 
    year = "2024"
}

@ARTICLE{Moroni2005,
    author = "Daniele Moroni AND Pieter Rein ten Wolde AND Peter G. Bolhuis", 
    title = "Interplay between Structure and Size in a Critical Crystal Nucleus", 
    journal = "Phys. Rev. Lett.",
    volume = "94", 
    pages = "235703", 
    year = "2005"
}

@ARTICLE{Tipeev2018,
    author = "A. O. Tipeev AND E. D. Zanotto AND J. P. Rino", 
    title = "Diffusivity, Interfacial Free Energy, and Crystal Nucleation in a Supercooled Lennard-Jones Liquid", 
    journal = "J. Phys. Chem. C",
    volume = "122", 
    pages = "28884", 
    year = "2018"
}

@ARTICLE{cheng2017.2,
    author = "Bingqing Cheng AND Gareth A. Tribello AND Michele Ceriotti1", 
    title = "The Gibbs free energy of homogeneous nucleation: From atomistic nuclei to the planar limit", 
    journal = "J. Chem. Phys.",
    volume = "147", 
    pages = "104707", 
    year = "2017"
}

@ARTICLE{Auer2001,
    author = "Stefan Auer AND Daan Frenkel", 
    title = "Prediction of absolute crystal-nucleation rate in hard-sphere colloids", 
    journal = "Nature",
    volume = "409", 
    pages = "1020–1023", 
    year = "2001"
}

@ARTICLE{tenWolde1996,
    author = "Pieter Rein ten Wolde AND Maria J. Ruiz‐Montero AND Daan Frenkel", 
    title = "Numerical calculation of the rate of crystal nucleation in a Lennard‐Jones system at moderate undercooling", 
    journal = "J. Chem. Phys.",
    volume = "104", 
    pages = "9932–9947", 
    year = "1996"
}

@ARTICLE{Gispen2023,
    author = "Willem Gispen AND Marjolein Dijkstra", 
    title = "Brute-force nucleation rates of hard spheres compared with rare-event methods and classical nucleation theory", 
    journal = "J. Chem. Phys.",
    volume = "159", 
    pages = "086101", 
    year = "2023"
}

@ARTICLE{Filion2010,
    author = "L. Filion AND M. Hermes AND R. Ni AND M. Dijkstra", 
    title = "Crystal nucleation of hard spheres using molecular dynamics, umbrella sampling, and forward flux sampling: A comparison of simulation techniques", 
    journal = "J. Chem. Phys.",
    volume = "133", 
    pages = "244115", 
    year = "2010"
}

@ARTICLE{baidakov2012,
    author = "Vladimir G. Baidakov AND Azat O. Tipeev", 
    title = "Crystal nucleation and the solid–liquid interfacial free energy", 
    journal = "J. Chem. Phys.",
    volume = "136", 
    pages = "074510", 
    year = "2012"
}

@ARTICLE{Leyssale2007,
    author = "Jean-Marc Leyssale AND Jérôme Delhommelle AND Claude Millot", 
    title = "Hit and miss of classical nucleation theory as revealed by a molecular simulation study of crystal nucleation in supercooled sulfur hexafluoride", 
    journal = "J. Chem. Phys.",
    volume = "127", 
    pages = "044504", 
    year = "2007"
}

@ARTICLE{Auer2004,
    author = "S. Auer; D. Frenkel", 
    title = "Numerical prediction of absolute crystallization rates in hard-sphere colloids", 
    journal = "J. Chem. Phys.",
    volume = "120", 
    pages = "3015–3029", 
    year = "2004"
}

@ARTICLE{bai2008,
    author = "Xian-Ming Bai and Mo Li", 
    title = "Comparing crystal–melt interfacial free energies through homogeneous nucleation rates", 
    journal = "J. Phys.: Condens. Matter",
    volume = "20", 
    pages = "375103", 
    year = "2008"
}

@ARTICLE{bai2006,
    author = "Xian-Ming Bai AND Mo Li", 
    title = "Calculation of solid-liquid interfacial free energy: A classical nucleation theory based approach", 
    journal = "J. Chem. Phys.",
    volume = "124", 
    pages = "124707", 
    year = "2006"
}

@ARTICLE{Chen2025,
    author = "Bin Chen", 
    title = "Classical Nucleation Theory and Tolman Equation in Cluster Thermodynamics: How Small Can They Truly Apply?", 
    journal = "J. Phys. Chem. A",
    volume = "129", 
    pages = "6018–6023", 
    year = "2025"
}

@ARTICLE{Iida2023,
    author = "Yuya Iida AND Tatsumasa Hiratsuka AND Minoru T. Miyahara AND Satoshi Watanabe", 
    title = "Mechanism of Nucleation Pathway Selection in Binary Lennard-Jones Solution: A Combined Study of Molecular Dynamics Simulation and Free Energy Analysis", 
    journal = "J. Phys. Chem. B",
    volume = "127", 
    issue = "15",
    pages = "3524–3533", 
    year = "2023"
}

@ARTICLE{Alexandrov2023,
    author = "Dmitri V. Alexandrov AND Eugenya V. Makoveeva", 
    title = "Two-step nucleation and crystal growth in a metastable solution", 
    journal = "J. Appl. Phys.",
    volume = "134", 
    pages = "234701", 
    year = "2023"
}

@ARTICLE{Gispen2025,
    author = "Willem Gispen AND Peter G. Bolhuis AND Marjolein Dijkstra", 
    title = "Kinetic phase diagram for two-step nucleation in colloid–polymer mixtures", 
    journal = "J. Chem. Phys.",
    volume = "162", 
    pages = "134901", 
    year = "2025"
}

@ARTICLE{Leyssale2005,
    author = "Jean-Marc Leyssale AND Jerome Delhommelle AND Claude Millot", 
    title = "Atomistic simulation of the homogeneous nucleation and of the growth of N2 crystallites", 
    journal = "J. Chem. Phys.",
    volume = "122", 
    pages = "104510", 
    year = "2005"
}

@ARTICLE{Gispen2025.2,
    author = "Willem Gispen AND Alberto Pérez de Alba Ortíz AND Marjolein Dijkstra", 
    title = "The bcc coating of Lennard-Jones crystal nuclei vanishes with a change of local structure detection algorithm", 
    journal = "J. Chem. Phys.",
    volume = "162", 
    pages = "024502", 
    year = "2025"
}

@ARTICLE{Ghiringhelli2007,
    author = "L. M. Ghiringhelli AND C. Valeriani AND E. J. Meijer AND D. Frenkel", 
    title = "Local Structure of Liquid Carbon Controls Diamond Nucleation", 
    journal = "Phys. Rev. Lett.",
    volume = "99", 
    pages = "055702", 
    year = "2007"
}

@ARTICLE{caplan2020,
    author = "M. E. Caplan AND C. J. Horowitz AND A. Cumming", 
    title = "Neon Cluster Formation and Phase Separation during White Dwarf Cooling", 
    journal = "ApJL",
    volume = "902", 
    pages = "L44", 
    year = "2020"
}

@ARTICLE{Lechner2008,
    author = "Wolfgang Lechner AND Christoph Dellago", 
    title = "Accurate determination of crystal structures based on averaged local bond order parameters", 
    journal = "J. Chem. Phys.",
    volume = "129", 
    pages = "114707", 
    year = "2008"
}

@ARTICLE{HamaguchiTriplePoint1997,
    author = "S. Hamaguchi AND R. T. Farouki AND D. H. E. Dubin", 
    title = "Triple point of {Y}ukawa systems", 
    journal = "Phys. Rev. E",
    volume = "56", 
    pages = "4671", 
    year = "1997"
}

@ARTICLE{wedekind2007,
    author = "Jan Wedekind AND Reinhard Strey AND David Reguera", 
    title = "New method to analyze simulations of activated processes", 
    journal = "J. Chem. Phys.",
    volume = "126", 
    pages = "134103", 
    year = "2007"
}

@ARTICLE{Nicholson2016,
    author = "David A. Nicholson AND Gregory C. Rutledge", 
    title = "Analysis of nucleation using mean first-passage time data from molecular dynamics simulation", 
    journal = "J. Chem. Phys.",
    volume = "144", 
    pages = "134105", 
    year = "2016"
}

@ARTICLE{espinosa2016,
    author = "Jorge R. Espinosa AND Carlos Vega AND Chantal Valeriani AND Eduardo Sanz", 
    title = "Seeding approach to crystal nucleation", 
    journal = "J. Chem. Phys.",
    volume = "144", 
    pages = "034501", 
    year = "2016"
}

@ARTICLE{Zimmerman2018,
    author = "Nils. E. R. Zimmermann AND Bart Vorselaars AND Jorge R. Espinosa AND David Quigley AND William R. Smith AND Eduardo Sanz AND Carlos Vega AND Baron Peters", 
    title = "NaCl nucleation from brine in seeded simulations: Sources of uncertainty in rate estimates", 
    journal = "J. Chem. Phys.",
    volume = "148", 
    pages = "222838", 
    year = "2018"
}

@ARTICLE{Farouki1994,
    author = "S. Hamaguchi AND R. T. Farouki", 
    title = "Thermodynamics of strongly‐coupled {Y}ukawa systems near the one‐component‐plasma limit. I. Derivation of the excess energy ", 
    journal = "J. Chem. Phys.",
    volume = "101", 
    pages = "9876–9884", 
    year = "1994"
}

\end{document}



\title{Supplementary Material for ``Deriving Reliable Nucleation Rates from Metadynamics Simulations: Application to Yukawa Fluids''} 



\author{B.~Arnold}
\affiliation{Laboratory for Laser Energetics, University of Rochester, Rochester, New York 14623, USA}
\affiliation{Department of Physics and Astronomy, University of Rochester, Rochester, New York 14611, USA }

\author{J.~Daligault}
\author{D.~Saumon}
\affiliation{Los Alamos National Laboratory, P.O. Box 1663, Los Alamos, New Mexico 87545, USA}

\author{S. X.~Hu}
\affiliation{Laboratory for Laser Energetics, University of Rochester, Rochester, New York 14623, USA}
\affiliation{Department of Physics and Astronomy, University of Rochester, Rochester, New York 14611, USA }
\affiliation{Department of Mechanical Engineering, University of Rochester, Rochester, New York 14611, USA }


\date{\today}

\pacs{}

\maketitle 

\section{Metadynamics Details}

We outline here how all parameters that must be specified to execute a metaD simulation were chosen.

The metaD bias potential is constructed by the deposition of many repulsive Gaussians. For the subspace- (globally-) biased simulations, these Gaussians have a width of 0.05 (0.02) and an initial height of roughly 1 (10) $k_BT$\cite{Branduardi2012}. One of these Gaussians is deposited every 1000 MD timesteps and all collective variables are output every 100 timesteps to construct as histogram as in Equation (7) of the main text. Throughout the simulation, the height of the Gaussians are gradually decreased to avoid pushing the simulation into unphysical configuration and improve convergence of the free energy \cite{Barducci2008, Barducci2011}. This is achieved with well-tempered metadynamics in PLUMED \cite{PLUMED} with a bias factor of 10 (200).

The size of the simulation must be large enough that a critical cluster can fit inside the box, but not so large that there are convergence issues. If, for example, one attempted a metaD calculation of the nucleation barrier at $\Theta = 0.85$, the critical cluster may contain $1000$ particles \cite{Arnold2025}. The simulation must contain many more particles than this so that configurations containing a critical cluster surrounded by liquid may be generated. If there are too few particles in the simulation, the nucleation-and-growth mechanism of the phase transition may not be captured \cite{Santos2022}, or the critical cluster may be elongated, reaching across periodic boundaries rather than being spherical \cite{Quiggley2009}. Both of these issues would suppress the calculated free energy barrier. Avoiding these issues requires a large number of particles in the simulations. However, when there are many thousands of particles in the simulation, it becomes improbable for a configuration to appear that contains no clusters. This is important because the overall normalization of the CNT free energy surface is determined by $F(0)$. When there are too many particles, the convergence of the simulation slows down because fewer of these cluster-free configurations can be sampled, and because calculating the collective variables becomes expensive. To balance these competing requirements between needing many particles to capture the nucleation mechanism and needing few particles to achieve convergence, we chose $N_\text{in}\approx N^*$ and $N_\text{tot}\approx10N^*$. With these requirements, the simulations range from $N_\text{in}\approx 200$ and $N_\text{tot}= 2662$ at $\Theta = 0.725$ to $N_\text{in}\approx 700$ and $N_\text{tot}= 5488$ at $\Theta=0.85$.

During the simulation, we calculate $N_\text{l}$, the number of particles in the largest crystalline cluster. Every particle is classified as either liquid if its $Q_6$ value is less than a threshold value,  or solid if $Q_6$ is greater than the threshold. We choose a threshold value of $Q_6$ such that in independent simulations of a bulk liquid and a bulk solid, total number of misidentified particles is minimized. Then, if two particles are both solid and within each other's first coordination sphere, they are considered to be part of the same cluster. PLUMED carries out this computation efficiently \cite{Tribello2017}, so we calculate the size of the largest cluster in the simulation on-the-fly.

\section{Wall Bias}
\label{sec:walls}

\subsection{The Need for Walls}
There are two major challenges to running metadynamics with a global order parameter. First, the barrier is very small relative to the free energy difference of the two wells. Most of the free energy barriers calculated in this paper are between $10$ and $50k_BT$. Since we are simulating systems below the melt temperature, the solid sits at the bottom of a free energy well approximately $N_\text{tot}\Delta \mu$ below the free energy of the liquid. for most of our simulations, this is of order $1000 k_BT$. Filling this deep free energy well in a computationally tractable metaD simulation requires constructing $V_\text{b}(Q_6)$ from large Gaussians. These large Gaussians will be bigger than the barrier of interest, making it almost impossible to accurately resolve the barrier. This well is also much broader than the extent of the barrier in $Q_6$-space. If the system is allowed to explore the whole solid well, it will spend very little time sampling the barrier. These issue can be partially resolved by adding a ``wall" that prevents the simulation from sampling highly crystallized configurations, focusing computational resources on sampling the free energy barrier. 

In the subspace-biased metaD, the ``outside'' liquid region must be forced to remain liquid under supercooled conditions. We accomplish this will a wall potential as well.

\subsection{Wall Implementation}

PLUMED allows the user to prevent the simulation from reaching certain values of collective variables by applying a harmonic wall constraint of the form
\begin{equation}
    V_{\text{w}}(Q) = \begin{cases}
        \epsilon\left(Q-q\right)^2 , & \text{if $Q>q$}\\
        0, & \text{otherwise}
        \end{cases}
\end{equation}
in addition to the metaD bias \cite{PLUMED}. This potential penalizes configurations with an order parameter $Q$ greater than the cutoff $q$. We use wall strengths $\epsilon \approx 10^5 k_BT$ for all simulations.

For the global simulations, we place the walls at $Q_\text{tot}=0.4$, and penalize configurations with overly large crystallinity. This corresponds to preventing a single cluster from getting too large, or to preventing many separate clusters from forming. This results in the metaD Hamiltonian 
\begin{equation}
    H = H_0 + V_\text{mD}(Q_\text{tot}) + V_\text{w}(Q_\text{tot}).
\end{equation}

The subspace-biased simulations require a constraint to prevent the ``outside'' region from solidifying at all. This requires a wall bias on only $Q_\text{out}$ with the stricter wall location $Q_\text{out}=0.25$. This gives the subspace-biased simulations the final metaD Hamiltonian 
\begin{equation}
    H = H_0 + V_\text{mD}(Q_\text{in}) + V_\text{w}(Q_\text{out}).
\end{equation} 

When we reweight simulations to estimate the unbiased free energies using balanced exponential reweighting \cite{Schafer2020}, we compensate for both the metaD bias potential and the wall potential.

\subsection{Side Effects and Limitations}

First, the outside wall interferes with the large-$Q_\text{in}$ part of the subspace-biased free energy surface. In Figure 4C, the crystal has grown to be larger than the biased region. Since it is growing into the outside region which is constrained to be liquid, this configuration has a large energy coming from the $V_{\text{w}}$ part of the metaD Hamiltonian. This leads to the nonphysical increase in free energy for large crystals, as is visible in Figure 7. When the biased region is large enough to accommodate the entire critical cluster, the peak of the barrier is still discernible.

The outside wall \emph{penalizes} configurations with clusters outside the biased region, but it does not \emph{prevent} such configurations from occuring. When temperature is very low (below $\Theta = 0.725$),  the thermodynamic driving force towards nucleation is too strong for the wall to prevent cluster formation. resolving the full free energy barrier requires sampling configurations with no clusters. When the wall is insufficient to stop the perpetual presence of clusters in the outside region below $\Theta=0.725$, subspace biasing can no longer resolve the nucleation barrier. 


\section{uncertainty estimates}
There is some ambiguity in the range of cluster sizes $N$ that should be included in the fits of section VA. Clearly, the large-$N$ region where the metaD FES increases due to the influence of the wall potential must be discarded. These discarded portions are shown in gray in Fig. (7). We estimate uncertainties on the fit values of $\gamma_\infty$ and $\delta$  by doing many fits to one free energy surface at a time while varying the range of $N$ over which the fit is performed. The lower end of the range is between $N=50$ and $100$ while the upper end of the range is between the peak and the first minimum.  (We perform one fit on the $\Theta=0.825$ FES over all data $50\leq N \leq 1000$, another over all data $75\leq N \leq 900$, etc.) This amounts to roughly 500 fits performed for each $\kappa$, each giving a different set of parameters $\gamma_\infty$ and $\delta$, cummulatively providing the uncertainty bands in Figs. (7), (8), and (10).


\bibliography{YOCPmetaD}